\documentclass[12pt]{article}
\RequirePackage[OT1]{fontenc}
\usepackage{amsthm,amsmath,amssymb,amsfonts,bm,bbm,latexsym,eucal}

\usepackage{graphicx,graphics,xcolor,color,tikz}
\usepackage{subcaption,adjustbox,wrapfig,pdfpages}
\usepackage{epsfig,epstopdf}
\usepackage{pstricks, pst-node, psfrag}

\usepackage{booktabs,multirow,makecell,longtable,arydshln}

\usepackage{algorithm,algpseudocode}

\usepackage{natbib,breakcites}
\RequirePackage[colorlinks,citecolor=blue,urlcolor=blue]{hyperref}

\usepackage{authblk}
\usepackage[toc,page]{appendix}
\usepackage{fancyhdr,sectsty}

\usepackage[normalem]{ulem}
\usepackage{pifont,textcmds,textcomp}
\usepackage{setspace}
\usepackage{verbatim,enumerate,enumitem}
\usepackage{listings}

\usepackage{etoolbox}
\usepackage{listofitems}
\usepackage{float}
\usepackage{rotating, lscape}
\usepackage{relsize}
\usepackage[usestackEOL]{stackengine}
\usepackage{lipsum}
\usepackage{academicons}
\usepackage{url}
\usepackage{wasysym}

\allowdisplaybreaks

\definecolor{lb}{RGB}{44,139,183}

\def\frac#1#2{{\textstyle{#1\over#2}}}

\DeclareSymbolFont{AMSb}{U}{msb}{m}{n}
\DeclareMathSymbol{\Natural}{\mathbin}{AMSb}{"4E}
\DeclareMathSymbol{\Integer}{\mathbin}{AMSb}{"5A}
\DeclareMathSymbol{\Real}{\mathbin}{AMSb}{"52}
\DeclareMathSymbol{\Rational}{\mathbin}{AMSb}{"51}
\DeclareMathSymbol{\Imaginary}{\mathbin}{AMSb}{"49}
\DeclareMathSymbol{\Complex}{\mathbin}{AMSb}{"43} 
\DeclareMathSymbol{\Disk}{\mathbin}{AMSb}{"44} 
\def\bi{\begin{itemize}}
\def\ei{\end{itemize}}
\def\bd{\begin{description}}
\def\ed{\end{description}}
\def\ben{\begin{enumerate}}
\def\een{\end{enumerate}}
\def\bv{\begin{verbatim}}
\def\ev{\end{verbatim}}

\def\bar#1{{\overline{#1}}}
\def\hat#1{{\widehat{#1}}}

\def\2to{{\ {\buildrel 2\over \longrightarrow}\ }}

\def\I1ton{{$I_1,\ldots,I_n$}}
\def\X1ton{{$X_1,\ldots,X_n$}}
\def\Y1ton{{$Y_1,\ldots,Y_n$}}
\def\Z1ton{{$Z_1,\ldots,Z_n$}}
\def\R1ton{{$R_1,\ldots,R_n$}}
\def\e1ton{{$e_1,\ldots,e_n$}}
\def\t1ton{{$t_1,\ldots,t_n$}}
\def\x1ton{{$x_1,\ldots,x_n$}}
\def\y1ton{{$y_1,\ldots,y_n$}}
\def\z1ton{{$z_1,\ldots,z_n$}}

\newcommand{\bfs}{\mathbf{s}}

\newcommand{\pkg}[1]{\texttt{#1}}
\title{Modeling nonstationary spatial processes with normalizing flows}

\author[1]{Pratik Nag}
\author[2,1]{Andrew Zammit-Mangion}
\author[3]{Ying Sun}

\affil[1]{School of Mathematics and Physics, University of Wollongong, Wollongong, Australia
\vspace{0.1in}}

\affil[2]{School of Mathematics and Statistics, University of New South Wales, Sydney, Australia
\vspace{0.1in}}

\affil[3]{CEMSE Division, Statistics Program, King Abdullah University of Science and Technology, Thuwal, Saudi Arabia}

\date{}

\begin{document}

\maketitle


\begin{center} 
{\large{\bf Abstract}}
\bi
\singlespacing
Nonstationary spatial processes can often be represented as stationary processes on a warped spatial domain. Selecting an appropriate spatial warping function for a given application is often difficult and, as a result of this, warping methods have largely been limited to two-dimensional spatial domains. In this paper, we introduce a novel approach to modeling nonstationary, anisotropic spatial processes using neural autoregressive flows (NAFs), a class of invertible mappings capable of generating complex, high-dimensional warpings. Through simulation studies we demonstrate that a NAF-based model has greater representational capacity than other commonly used spatial process models. We apply our proposed modeling framework to a subset of the 3D Argo Floats dataset, highlighting the utility of our framework in real-world applications.
\ei
\end{center}

\section{Introduction} 
\label{sec:introduction}
Nonstationary spatial process models are widely used across various fields such as environmental science, geostatistics, epidemiology, and remote sensing, to analyze spatial phenomena that exhibit spatially varying patterns \citep{banerjee2003hierarchical, higdon2022non}. Unlike stationary process models, whose statistical properties are invariant across space, nonstationary process models have spatially varying characteristics, enabling them to capture complex spatial dependencies. Traditionally, spatial processes have been modeled using Gaussian process with stationary covariance functions \citep{cressie1993statistics, stein2005space}. However, these models often fail to adequately capture the spatial variability often seen in real-world datasets, potentially leading to biased inference and inaccurate predictions. To address this limitation, there has been considerable effort in the last few decades to construct classes of valid nonstationary spatial process models.

Techniques for modeling covariance nonstationarity include approaches based on spatially varying parameters, spatially varying kernels, and hierarchical models \citep{risser2016review}. Among these, the nonstationary Mat\'ern covariance function \citep{paciorek2006spatial} is ubiquitous. 

This covariance function provides a flexible framework for capturing spatial dependencies that themselves change in space, but requires tailored construction for the problem at hand. Another popular method is the stochastic partial differential equation approach introduced by \cite{lindgren2011explicit}.

While these covariance-based approaches offer powerful tools for analyzing spatial data, they make various modeling assumptions that can limit their flexibility. In response to some of their limitations, recent years have seen increased use of neural networks for capturing spatial nonstationarity. 
One of these approaches, coined ``DeepKriging" \citep{chen2020deepkriging}  passes the output of a conventional fixed rank kriging model \citep{cressie2008fixed} through a deep neural network; the nonlinearities in the network induce nonstationarity. This framework has been adapted to use convolutional neural networks by \cite{wang2024spatialdeepconvolutionalneural}, and extended to the spatio-temporal setting by \cite{nag2023spatio}.

An alternative approach to modeling covariance nonstationarity, which lends itself to neural networks, involves warping the spatial domain. Consider a spatial process \( Y(\cdot) \) defined on a spatial domain \( D \), where \( \text{var}(Y(\bfs)) < \infty \) for \( \bfs \in D \). The concept put forth by \citet{sampson1992nonparametric} proposes transforming the domain \( D \) via a mapping \( f : D \rightarrow D_1 \), such that the process has a stationary and isotropic covariance structure on \( D_1 \). In their seminal work, the mapping \( f(\cdot) \) was implemented using multidimensional scaling (MDS) and thin-plate splines. Since then, various alternative methods have been proposed. For example, \citet{smith1996estimating} suggested modeling the mapping \( f(\cdot) \) as a sum of radial basis functions derived from thin-plate splines, with basis-function coefficients estimated through a likelihood-based technique. \cite{perrin1999modelling} employed compositions of radial basis function mappings to construct \( f(\cdot) \). Other examples of spatial warping modeling approaches can be found in \cite{schmidt2003bayesian, castro2013state}; and \citet{ghulam_warping2021}.

A relatively new approach to model the deformation expresses \( f(\cdot) \) as a composition of simple functions. When warping space, it is important to ensure injectivity of \( f(\cdot)\) to avoid issues related to space folding (see, e.g.,  \cite{sampson1992nonparametric} and \cite{schmidt2003bayesian}). To this end, \citet{zammit2021deep} proposed a deep compositional spatial model, which builds on the premise that a map created by composing multiple injective maps is itself injective. Within this framework, \( f(\cdot) \) is constructed using injective radial basis functions, M{\"o}bius transformations, and monotonic axial warping units. These warping functions need to be chosen by the user, and a poor choice may constrain the type of heterogeneity being modeled. 

In this paper, we explore the use of a special class of autoregressive flows \citep[AFs, ][]{kobyzev2020normalizing}, called neural autoregressive flows \citep[NAFs, ][]{huang2018neural} to construct the injective mapping $f(\cdot)$. NAFs, described in Section \ref{sec:overview}, are injective. Although one needs to construct a neural network architecture to use NAFs, we see from our experiments that even a simple architecture can be used to generate a wide class of flexible nonstationary spatial process models. Our framework thus frees up the user from having to choose the warping units. Further, unlike the deep compositional spatial model of \citet{zammit2021deep},  it can also be used for higher-dimensional data with practically no change to the underlying architecture and software. This makes our framework applicable to a wide range of scenarios.

In Section \ref{sec:overview}, we present our new framework, and outline how it improves over existing work. We present two simulation experiments in Section \ref{sec:2_dim} that demonstrate the utility and benefits of our approach for modeling nonstationarity. Section \ref{sec:real_data} applies the methodology to the Argo float data, which is three-dimensional. Finally, Section \ref{sec:discussion} reflects on the key features of our proposed model, and outlines potential directions for future research.

\section{Overview}
\label{sec:overview}

Consider the real valued spatial field \{$Y(\mathbf{s}), \mathbf{s}\in D\}, D \subseteq \mathbb{R}^{d}$, and associated observations $\mathbf{Z} \equiv (Z_1,Z_2,\ldots,Z_N)'$ at $N$ spatial locations $\{\bfs_1,\ldots,\bfs_N\} \subset D$. We assume that the observations are noisy measurements of the process; specifically:
\begin{equation}
        Z_i = Y(\bfs_i) + \epsilon_i, \quad \bfs_i \in D, \quad i = 1,\ldots, N,
        \label{eq:obs}
    \end{equation}
where $\epsilon_i \sim \text{Gau}(0,\sigma^2_\epsilon)$, for $i = 1,\dots,N$, is independent and identically distributed Gaussian measurement error with variance $\sigma^2_\epsilon$. Here, we consider the case where the latent process $Y(\cdot)$ is a zero-mean spatial Gaussian process with covariance function $C(\bfs,\mathbf{u}), \ \bfs,\mathbf{u} \in D$, that does not admit a known parametric form. The model for $Y(\cdot)$ is important when making predictions with spatially-sparse data, or data with a low signal-to-noise ratio.

Several approaches have been proposed for constructing $C(\cdot,\cdot)$ such that it can encode covariance nonstationarity, nonseparability, and asymmetry, in a parsimonious way. Here, we focus on the deformation method, first proposed by \cite{sampson1992nonparametric}, which models a simple stationary covariance function on a warped spatial domain. In this setting, the parameters $(\boldsymbol{\vartheta}', \boldsymbol{\phi}')'$ comprise those of a warping function, which we denote by $\boldsymbol{\vartheta}$, and those that parameterize the stationary covariance function on the warped domain, which we denote by $\boldsymbol{\phi}$. The challenge with this method is to fit a suitable warping function.

\cite{zammit2021deep} propose using a collection of simple warping units to model the warping. Their deep compositional spatial model only involves a few unknown parameters, and it can be efficiently implemented using GPUs and common deep learning software such as \pkg{Tensorflow}. Unlike several other warping models, the deep compositional spatial model can be fitted using a single spatial replicate. However there are two drawbacks of the model proposed by \cite{zammit2021deep}. First, the individual warping units need to be chosen by the user, and only a limited number of warpings are available. Second, the proposed warping functions are only suitable for one- and two-dimensional spatial process models. In the following sections, we introduce a method based on NAFs, which we later show are capable of modeling flexible warpings in three-and-higher dimensions.

\subsection{Auto-regressive flows}\label{sec:autoregressive_flows}

An AF \citep[][]{papamakarios2017masked} is a special case of a normalizing flow \citep[NF, ][]{kobyzev2020normalizing}, which is a flexible invertible function. We can view the AF as an increasing triangular map \(\mathbf{T}(\cdot)\) that can be parameterized in various ways \cite[see, e.g., ][]{dinh2015nice, germain2015made}. Let $\bfs \equiv (s_1,\ldots,s_d)' \in D$ denote a spatial coordinate in our geographic domain $D$. The triangular map $\mathbf{T}(\bfs; \boldsymbol{\vartheta}) \equiv (T^{(1)}(s_1; \boldsymbol{\vartheta}_1),\dots,T^{(d)}(s_1,\dots,s_d; \boldsymbol{\vartheta}_d))', \bfs \in D,$ is given by 
\begin{align}
    T^{(1)}({s}_1; \boldsymbol{\vartheta}_1) &= {S}^{(1)}({s}_1;\boldsymbol{\gamma}_1(\bm{\vartheta}_{1})), \\
    T^{(k)}({s}_1,\dots,{s}_k; \boldsymbol{\vartheta}_k) &= {S}^{(k)}({s}_k;\bm{\gamma}_{k}({s}_1,\dots,{s}_{k-1};\bm{\vartheta}_{k})), \ k = 2,\ldots,d,
\end{align}
where $\boldsymbol{\vartheta} \equiv (\boldsymbol{\vartheta}_1',\dots,\boldsymbol{\vartheta}_d')'$ are the parameters of the triangular map; $\bm{\gamma}_{1}(\bm{\vartheta}_{1})$ is a fixed transformation of $\bm{\vartheta}_{1}$ (see Section~\ref{sec:masking}); $\bm{\gamma}_{k}({s}_1,\dots,{s}_{k-1};\bm{\vartheta}_{k})$ for $k = 2,\dots,d,$ is a flexible parametric function, which in this work we construct using a neural network, of the first $k-1$ components of $\bfs \in D$, with parameters $\bm{\vartheta}_{k}$; and $S^{(k)}(\cdot)$ is a monotonic function. Therefore, each component of $\mathbf{T}(\cdot)$  except the first, involves a neural network that takes as input $({s}_1,\dots,{s}_{k-1})'$ and that outputs parameters used to construct ${S}^{(k)}(\cdot)$. That is, for $k \ge 2$, $\bm{\gamma}_{k}:\mathbb{R}^{k-1} \rightarrow \mathbb{R}^{m_k}$, where $m_k$ is the number of parameters that parameterize ${S}^{(k)}$.  For ease of exposition, for the remainder of this section and in Section~\ref{sec:naf} we omit the dependence of $\bm{\gamma}_{k}(\cdot)$ on its inputs and parameters, and simply denote it as $\bm{\gamma}_{k}$.

It remains to define $S^{(k)}$, for $k = 1,\dots, d$. One option for $S^{(k)}$ is
\begin{equation}
    S^{(k)}(s_k;\bm{\gamma}_{k}) = \sigma(\gamma_{k}^{(2)})s_k + (1 - \sigma(\gamma_{k}^{(2)}))\gamma_{k}^{(1)}, \quad k = 1,\dots,d,
    \label{eq:iaf}
\end{equation}
where $\sigma(\cdot)$ is the sigmoid function. Since $\sigma(\cdot) > 0$, $S^{(k)}(s_k;\bm{\gamma}_{k})$ is a weighted average of $s_k$ and $\gamma_{k}^{(1)}$. Clearly, $S^{(k)}(\cdot)$ is a monotonic function of its input, and it can be shown that the resulting triangular map $\mathbf{T}(\cdot)$ is invertible. The invertibility ensures that the warping function does not fold space onto itself. Equation \eqref{eq:iaf} is simple in form, with $m_k = 2$ for $k = 1,\ldots, d$. In practice, we require more flexibility in our transformation. Here, we achieve this flexibility using NAFs. 

\subsection{Neural auto-regressive flows}\label{sec:naf}
Neural autoregressive flows are autoregressive flows where each $S^{(k)}(\cdot), k = 1,\dots, d$, is itself modeled using an (injective) neural network \citep{huang2018neural}. There are several types of NAFs, and one of the most used is the deep sigmoidal flow (DSF). In the DSF, each $S^{(k)}(\cdot)$ is modeled using a single layer neural network each with $m_k = 3M \text{ parameters, where } M \geq 1$. In the DSF, $S^{(k)}(\cdot)$ has the form
\begin{equation}
    {S}^{(k)}({s}_k;\bm{\gamma}_{k}) = \sigma^{-1}\left( \mathbf{w}_{k}'  \sigma(\mathbf{a}_{k} {s}_k + \mathbf{b}_{k})\right), \ k = 1,\ldots,d.
    \label{eq:dsf}
\end{equation}
In \eqref{eq:dsf}, $\sigma^{-1}(\cdot)$ is the logit function that is applied elementwise to its vector input, and the parameters $\bm{\gamma}_k \equiv (\mathbf{w}_{k}',\mathbf{a}_{k}',\mathbf{b}_{k}')'$ are the $3M$ neural network parameters, with $\mathbf{w}_k, \mathbf{a}_k$ and $\mathbf{b}_k$ each of length $M$ and with $\sum_{i=1}^{M}{w}_{k,i} = 1$. This construction ensures monotonicity of the function ${S}^{(k)}(\cdot)$ and invertibility of the multivariate map $\mathbf{T}(\cdot)$.

The multi-layer extension of the DSF is the deep dense sigmoidal flow (DDSF). Consider an $L$-layer DDSF where the output dimension of each layer is $M_l$, $l =1, \ldots, L$, with $M_L=1$, and the ``inner" dimension of each layer is $\tilde{M}_l$, $l = 1, \ldots, L$. In the DDSF, ${S}^{(k)}(\cdot)$ is constructed as follows:
 \vspace{1mm}
\begin{equation}
    \begin{aligned}
    \mathbf{h}^1_k(s_k;\bm{\gamma}_{k}) &= \sigma^{-1}\left( \mathbf{W}_{k}^{1}  \ \sigma\left(\mathbf{a}_{k}^{1} s_k + \mathbf{b}_{k}^{1}\right)\right), \\
    \mathbf{h}^{l}_k(s_k;\bm{\gamma}_{k}) &= \sigma^{-1}\left( \mathbf{W}_{k}^{l} \ \sigma\left(\mathbf{a}_{k}^{l} \odot (\mathbf{U}_{k}^{l} \ \mathbf{h}^{l-1}_k(s_k;\bm{\gamma}_{k})) + \mathbf{b}_{k}^{l}\right)\right), \quad l = 2,\ldots,L-1, \\
    {S}^{(k)}(s_k;\bm{\gamma}_{k}) &= {h}^L_k(s_k;\bm{\gamma}_{k}) = \sigma^{-1}\left( {\mathbf{w}_{k}^{L}}' \ \sigma\left(\mathbf{a}_{k}^{L} \odot (\mathbf{U}_{k}^{L} \ \mathbf{h}^{L-1}_k(s_k;\bm{\gamma}_{k})) + \mathbf{b}_{k}^{L}\right)\right), \\
\end{aligned}
\label{eq:ddsf}
\end{equation}
where $\mathbf{a}_k^l$ and $\mathbf{b}_k^l$ are $\tilde{M}_l$ vectors for $l = 1,\dots,L$,   $\mathbf{W}_{k}^{l} \text{ is an } M_l \times \tilde{M}_l$ matrix for $l = 1, \ldots, L-1$, and $\mathbf{U}_{k}^{l} \text{ is an } \tilde{M}_{l} \times M_{l-1}$ matrix for $l = 2, \ldots, L$. The quantity $\mathbf{w}_{k}^{L}$ in the final layer is of length $\tilde{M}_L$. To ensure monotonicity of $S^{(k)}(\cdot)$ the following constraints are imposed \citep{huang2018neural}:

\begin{equation}
  \begin{aligned}
    &\sum_{j=1}^{\tilde{M}_l}{W}_{k,ij}^l = 1  \quad i = 1, \ldots, {M}_l,~~ l = 1,\ldots L-1,\\
    &\sum_{j=1}^{M_{l-1}}{U}_{k,ij}^l = 1,  \quad i = 1, \ldots, \tilde{M}_l,~~ l = 2,\ldots L,\\
    &\sum_{j=1}^{\tilde{M}_L}{w}_{k,j}^L = 1, \quad 
    \end{aligned}
\end{equation}
where all the parameters except $\mathbf{b}_k^l$, for $l = 1,\dots, L$, are positive. In this case, $$\boldsymbol{\gamma}_k \equiv (\mathbf{a}_k^{1'},\dots,\mathbf{a}_k^{L'},\mathbf{b}_k^{1'},\dots,\mathbf{b}_k^{L'},\text{vec}(\mathbf{W}_k^{1})',\dots,\text{vec}(\mathbf{W}_k^{(L-1)})',\mathbf{w}_k^{L'},\text{vec}(\mathbf{U}_k^{2})',\dots,\text{vec}(\mathbf{U}_k^{L})')'.$$

Throughout this work we will employ a DDSF architecture for constructing the triangular map. We collect in $\boldsymbol{\gamma} \equiv (\boldsymbol{\gamma}_1',\dots,\boldsymbol{\gamma}_d')'$ all the parameters that parameterize this map, which are themselves outputs of neural networks. Recall that  $\boldsymbol{\gamma}_k$, which  dictates the transformation at the $k$th step in the flow, has $m_k$ parameters. We denote the total number of parameters in the transformation as $m = \sum_{k=1}^d m_k$. We implement the NAF using masking as discussed next.

\subsection{Binary masking}\label{sec:masking}

In this section, we re-introduce the dependence of the $m$-dimensional vector function $\boldsymbol{\gamma}(\bfs; \boldsymbol{\vartheta})$ on its input $\bfs \in D$ and parameters $\boldsymbol{\vartheta}$. Further, with a slight abuse of notation, we now assume that the $m_k$-dimensional vector function $\boldsymbol{\gamma}_k(\cdot\,; \boldsymbol{\vartheta}_k), k = 1,\dots, d$, takes the entire spatial vector $\bfs$ as input. In this section we describe how we can implement $\boldsymbol{\gamma}(\cdot)$ with a single neural network.

There are two main considerations when implementing a neural network for $\boldsymbol{\gamma}(\cdot)$: First, several of its components have either positivity or sum-to-one constraints, which we accommodate by using suitable activation functions in the final layer of the network (e.g., the exponential activation function guarantees positivity, while the softmax function guarantees a sum-to-one constraint). Second, to ensure invertibility, $\boldsymbol{\gamma}_1(\bfs; \boldsymbol{\vartheta}_1)$ has to be invariant to $\bfs$, while $\boldsymbol{\gamma}_k(\bfs; \boldsymbol{\vartheta}_k)$ for $k = 2,\dots,d$, must only depend on $(s_1,\dots,s_{k-1})'$. We meet this second requirement through the use of masks, using the approach proposed for the masked autoencoder for distribution estimation (MADE) network \citep{germain2015made}.

For ease of exposition, assume that we model $\boldsymbol{\gamma}_k(\cdot)$ using a feedforward neural network with no hidden layers.  In order to ensure $\boldsymbol{\gamma}_k(\bfs; \boldsymbol{\vartheta}_k), \bfs \in D,$ is invariant to $s_k,\dots,s_d$ for $k = 1,\dots, d$, one can use a binary mask $\mathbf{M}_k$ in the network as follows,
\begin{equation}
  \begin{aligned}\label{eq:1layergamma}
\bm{\gamma}_k(\bfs; \boldsymbol{\vartheta}_k) ={\mathbf{g}_k}\left(\boldsymbol{\omega}_k+\left(\mathbf{H}_k \odot \mathbf{M}_k\right) \bfs\right), \quad \bfs \in D, \ k = 1,\dots,d,
\end{aligned}  
\end{equation}
where $\boldsymbol{\vartheta}_k \equiv (\text{vec}(\mathbf{H}_k)', \boldsymbol{\omega}_k' )'$; $\mathbf{H}_k$ and $\boldsymbol{\omega}_k$ are parameters that need to be estimated; ${\mathbf{g}(\cdot)}$ is a vector of nonlinear activation functions ensuring that the individual constraints on the elements of $\boldsymbol{\gamma}_k(\cdot )$ are met; and where $\odot$ denotes element-wise multiplication.
The masking matrix $\mathbf{M}_k$ is given by
\begin{equation}
   M_{k,{ij}}= \begin{cases}1 & \text{ if } j < k \\ 0 & \text{ otherwise }\end{cases}, \quad k = 2, \ldots, d,
\end{equation}
for $i = 1,\dots, m_k$ and $j  = 1,\dots, d$. The fixed masks ensure that $\boldsymbol{\gamma}_1(\bfs; \boldsymbol{\vartheta}_1)$ is invariant to $\bfs$, and that $\boldsymbol{\gamma}_k(\bfs; \boldsymbol{\vartheta}_k)$ only depends on $s_1,\dots,s_{k-1}$ for $k = 2,\dots, d$. A neural-network architecture for $\boldsymbol{\gamma}(\cdot)$ is then obtained by simply stacking \eqref{eq:1layergamma} for $k = 1,\dots,d$. Specifically, $\bm{\gamma}(\bfs ; \boldsymbol{\vartheta}) ={\mathbf{g}}\left(\boldsymbol{\omega}+\left(\mathbf{H} \odot \mathbf{M}\right) \bfs\right), \bfs \in D$, where $\mathbf{g}(\cdot) \equiv (\mathbf{g}_1(\cdot)',\dots,\mathbf{g}_d(\cdot)')'$; $\boldsymbol{\omega} \equiv (\boldsymbol{\omega}_1',\dots,\boldsymbol{\omega}_d')'$; $\mathbf{H} \equiv (\mathbf{H}_1',\dots,\mathbf{H}_d')'$; and $\mathbf{M} \equiv (\mathbf{M}_1',\dots,\mathbf{M}_d')'$; and where the transformation parameters are all collected into $\boldsymbol{\vartheta} \equiv (\text{vec}(\mathbf{H})', \boldsymbol{\omega}')'$.

In practice, one or more hidden layers are added to \eqref{eq:1layergamma} to increase model flexibility while maintaining the autoregressive property. When implementing hidden layers, one needs to use masks at each layer (that could be sparse and generated randomly) and keep track of which components of $\bfs$ the output at each hidden layer depends on. The masking matrix at the output layer is then constructed such that the $k$th output only depends on those hidden states that themselves depend on $s_1,\dots,s_{k-1}$. For more details, see \citet{germain2015made}.

\subsection{Fitting the model on the warped domain}\label{sec:loss}

Define the warped domain $\mathcal{D} \equiv \{\mathbf{T}(\bfs; \boldsymbol{\vartheta}): \bfs \in D\}$, and denote $\tilde{\bfs} = \mathbf{T}(\bfs; \boldsymbol{\vartheta})$ for $\bfs \in D$. We model the nonstationary process $Y(\cdot)$ in \eqref{eq:obs} as a mean zero Gaussian process on $\mathcal{D}$ with stationary and isotropic covariance function $\widetilde{C}_{\boldsymbol{\phi}}(h)$ for $h \equiv \|\tilde\bfs_i - \tilde\bfs_j\|$, $\tilde\bfs_i, \tilde\bfs_j \in \mathcal{D}$, where ${\boldsymbol{\phi}}$ parameterizes the covariance function on $\mathcal{D}$.
Then, for $N$ observations defined in \eqref{eq:obs}, the log-likelihood function of $\boldsymbol{\vartheta}$,  $\boldsymbol{\phi}$ and $\sigma_\epsilon$ is
\begin{equation}
    \ell(\boldsymbol{\vartheta},\boldsymbol{\phi}, \sigma_\epsilon; \mathbf{Z}) =  - \frac{N}{2} \log(2\pi) - \frac{1}{2} \log|\mathbf{K}_{\boldsymbol{\vartheta},\boldsymbol{\phi}, \sigma_\epsilon}| -\frac{1}{2} \mathbf{Z}' \mathbf{K}^{-1}_{\boldsymbol{\vartheta},\boldsymbol{\phi}, \sigma_\epsilon} \mathbf{Z},
\end{equation}
where 
\begin{equation}
\mathbf{K}_{\boldsymbol{\vartheta},\boldsymbol{\phi},\sigma_\epsilon} = (\widetilde{C}_{\boldsymbol{\phi}}(\|\mathbf{T}(\bfs_i; \boldsymbol{\vartheta}) - \mathbf{T}(\bfs_j; \boldsymbol{\vartheta})\|):  i,j = 1,\ldots,N) + \sigma^2_\epsilon \mathbf{I}.\label{eq:warped_cov}
\end{equation}

We employ a two-stage optimization procedure for maximizing the log-likelihood function. Denote the starting values for the three  unknown quantities in the optimization procedure as $\boldsymbol{\vartheta}^{[0]}$, $\boldsymbol{\phi}^{[0]}$ and $\sigma_\epsilon^{[0]}$. In the first stage, we maximize \( \ell(\boldsymbol{\vartheta},\boldsymbol{\phi}^{[0]}, \sigma_\epsilon^{[0]}) \) with respect to $\boldsymbol{\vartheta}$, using $\boldsymbol{\vartheta}^{[0]}$ as the starting point for the gradient ascent, to yield $\boldsymbol{\vartheta}^{[1]}$.  In the second stage, we maximize \( \ell(\boldsymbol{\vartheta}^{[1]},\boldsymbol{\phi}, \sigma_\epsilon) \) with respect to $\boldsymbol{\phi}$ and $\sigma_\epsilon$, using $\boldsymbol{\phi}^{[0]}$ and $\sigma_\epsilon^{[0]}$ as the starting point for the gradient ascent, to yield $\boldsymbol{\phi}^{[1]}$ and $\sigma_\epsilon^{[1]}$. These two steps are iteratively repeated until the stationary-covariance-function parameters do not change substantially between two iterations, that is, until
\[
\|(\boldsymbol{\phi}^{[t]'},\sigma_\epsilon^{[t]})' - (\boldsymbol{\phi}^{[t-1]'},\sigma_\epsilon^{[t-1]})' \| < \psi,
\]
for some $t > 2$ and a small tolerance $\psi > 0$.  The reason we only assess convergence of the covariance function parameters is that we expect there to be several warpings for the same covariance-function parameters that yield the same likelihood; a way to deal with this unidentifiability issue, if it becomes of practical concern, is through \textit{homogenization}; see \citet{vu2022modeling} for more details. This block-coordinate ascent procedure, which converges to a local maximizer of the likelihood function, is summarized in Algorithm \ref{algo:param_estimation}. 

Once the maximum-likelihood estimates are found, spatial prediction proceeds through simple kriging. Specifically, denote the maximum likelihood estimates of $\boldsymbol{\vartheta}$, $\boldsymbol{\phi}$ and $\sigma_\epsilon$ as $\hat{\boldsymbol{\vartheta}}$, $\hat{\boldsymbol{\phi}}$ and $\hat{\sigma_\epsilon}$, respectively. The simple kriging predictor at a new location $\mathbf{s}_0$ is obtained by computing the covariance matrix \( \mathbf{K}_{\hat{\boldsymbol{\vartheta}},\hat{\boldsymbol{\phi}},\hat{\sigma_\epsilon}} \) using \eqref{eq:warped_cov}, and the covariance vector \( \mathbf{k}_{\hat{\boldsymbol{\vartheta}},\hat{\boldsymbol{\phi}}} = (\widetilde{C}_{\hat{\boldsymbol{\phi}}}(\|\mathbf{T}(\bfs_0; \hat{\boldsymbol{\vartheta}}) - \mathbf{T}(\bfs_j; \hat{\boldsymbol{\vartheta}})\|):  j = 1,\ldots,N)'\).  The simple kriging predictor is then

\begin{equation}
    \hat{Y}(\mathbf{s}_0) = \mathbf{k}_{\hat{\boldsymbol{\vartheta}},\hat{\boldsymbol{\phi}}}' \mathbf{K}_{\hat{\boldsymbol{\vartheta}},\hat{\boldsymbol{\phi}}, \hat{\sigma_\epsilon}}^{-1} \mathbf{Z}.
    \label{eq:gkriging}
\end{equation}

\noindent Prediction standard errors are obtained from the simple kriging variance, which is given by $\widetilde{C}_{\hat{\boldsymbol{\phi}}}(0) - \mathbf{k}_{\hat{\boldsymbol{\vartheta}},\hat{\boldsymbol{\phi}}}' \mathbf{K}_{\hat{\boldsymbol{\vartheta}},\hat{\boldsymbol{\phi}}, \hat{\sigma}_\epsilon}^{-1} \mathbf{k}_{\hat{\boldsymbol{\vartheta}},\hat{\boldsymbol{\phi}}}$. Deriving multiple predictions and prediction covariances is a straightforward extension of the one-at-a-time procedure for prediction described here.

The kriging predictor \eqref{eq:gkriging} uses the inverse of the full covariance matrix $\mathbf{K}_{\hat{\boldsymbol{\vartheta}},\hat{\boldsymbol{\phi}}, \hat{\sigma_\epsilon}}$. In practice, for large datasets, one can use a low-rank approximation \citep[e.g.,][]{Zammit_2021} or a nearest-neighbor approximation \citep{datta2016hierarchical} to reduce the computational cost. We provide a large-scale implementation of our method using a nearest-neighbor approximation \citep{katzfuss2021general} in the Supplementary Material, Section S.1.

\begin{algorithm}[t!]
    \caption{Parameter Estimation}
    \label{algo:param_estimation}
    \begin{algorithmic}[1]
    \Require Observations $\{\mathbf{s}_i, Z_i\}_{i=1}^N$, initial parameters $\boldsymbol{\vartheta}^{[0]}, \boldsymbol{\phi}^{[0]}$, $\sigma_\epsilon^{[0]}$, tolerance $\psi$
    \State Set $t \gets 0$
    \Repeat
    \State $t \gets t + 1$
    \State Find $\boldsymbol{\vartheta}^{[t]} = \arg\max_{\boldsymbol{\vartheta}} \ell(\boldsymbol{\vartheta}, \boldsymbol{\phi}^{[t-1]}, \sigma_\epsilon^{[t-1]})$
        \State Find $(\boldsymbol{\phi}^{[t]'}, \sigma_\epsilon^{[t]})' = \arg\max_{\boldsymbol{\phi},\sigma_\epsilon} \ell(\boldsymbol{\vartheta}^{[t]}, \boldsymbol{\phi}, \sigma_\epsilon)$
    \Until{$\|(\boldsymbol{\phi}^{[t]'},\sigma_\epsilon^{[t]})' - (\boldsymbol{\phi}^{[t-1]'},\sigma_\epsilon^{[t-1]})' \| < \psi$}
    \State Set $\hat{\boldsymbol{\vartheta}} \gets \boldsymbol{\vartheta}^{[t]},\ \hat{\boldsymbol{\phi}} \gets \boldsymbol{\phi}^{[t]},\ \hat{\sigma_\epsilon} \gets \sigma_\epsilon^{[t]}$
    \end{algorithmic}
\end{algorithm}

\subsection{Computational complexity}\label{sec:complexity}

In this section, we analyze the computational and memory complexity of Algorithm~\ref{algo:param_estimation} when the warping map $\mathbf{T}(\bfs;\boldsymbol{\vartheta})$ is parameterized via a NAF. At each outer iteration, the algorithm alternates between updating the warping parameters $\boldsymbol{\vartheta}$ and the stationary covariance parameters $\{\boldsymbol{\phi}, \sigma_\epsilon\}$. Both steps require repeated evaluations of the Gaussian process log-likelihood, while the former additionally involves repeated evaluations of the warping function.

Evaluation of the triangular map $\mathbf{T}(\bfs;\boldsymbol{\vartheta})$ at a single location $\bfs \in D$ involves two components: (i) the conditioner network that jointly produces $\bm{\gamma}(\bfs;\boldsymbol{\vartheta}) \equiv (\bm{\gamma}_1',\dots,\bm{\gamma}_d')'$, and (ii) the DDSF transformations. As described in Section~\ref{sec:masking}, the conditioner is implemented as a single MADE-style feedforward neural network with $L_c$ hidden layers and $H_c$ hidden units per layer, taking $\bfs \in \mathbb{R}^d$ as input and producing the full $m$-dimensional output $\bm{\gamma}(\bfs;\boldsymbol{\vartheta})$ in one forward pass, where $m = \sum_{k=1}^d m_k$. The autoregressive property is enforced by masking, which does not alter the cost. Since $H_c \gg M_l, \tilde{M}_l$ for all DDSF layers $l = 1,\dots,L$, the per-location evaluation cost is dominated by the conditioner, and we focus on its contribution.

The cost of a single forward pass through the conditioner is
\[
\mathcal{O}\big(d H_c + (L_c-1)H_c^2 + H_c m\big),
\]
corresponding to the input-to-hidden, hidden-to-hidden, and hidden-to-output transformations. The parameter count, and hence the memory required to store the conditioner, scales analogously. Evaluating the map at all $N$ observation locations therefore costs
\[
\mathcal{O}\!\left(N\big(d H_c + L_c H_c^2 + H_c m\big)\right).
\]

Next, we consider the computational complexity associated with evaluating the Gaussian process log-likelihood. Given the warped locations $\{\tilde{\bfs}_i\}_{i=1}^N$, the covariance matrix $\mathbf{K}_{\boldsymbol{\vartheta},\boldsymbol{\phi},\sigma_\epsilon}$ defined in \eqref{eq:warped_cov} is constructed by evaluating the stationary covariance function $\widetilde{C}_{\boldsymbol{\phi}}(\|\tilde{\bfs}_i - \tilde{\bfs}_j\|)$ for all pairs of locations. This requires $\mathcal{O}(N^2 d)$ operations. Evaluation of the log-likelihood itself involves computing the Cholesky factorization of the covariance matrix and performing the associated triangular solves, requiring $\mathcal{O}(N^3)$ floating-point operations and $\mathcal{O}(N^2)$ memory. This step dominates the computational cost for moderate to large sample sizes.

Let $I_{\vartheta}$ and $I_{\phi}$ denote the number of inner gradient-based optimization iterations used when updating $\boldsymbol{\vartheta}$ and $\{\boldsymbol{\phi},\sigma_\epsilon\}$, respectively. Each inner iteration of the $\boldsymbol{\vartheta}$-update requires re-evaluating the warping map at all $N$ locations, recomputing the covariance matrix, and performing a Cholesky factorization. Each inner iteration of the $\{\boldsymbol{\phi},\sigma_\epsilon\}$-update, for fixed $\boldsymbol{\vartheta}$, does not require re-evaluating the warping map, but still requires rebuilding the covariance matrix (as $\boldsymbol{\phi}$ changes) and performing a Cholesky factorization. The computational cost per outer iteration is therefore
\[
\mathcal{O}\!\left(
(I_{\vartheta}+I_{\phi})\, N^3
+
(I_{\vartheta}+I_{\phi})\, N^2 d
+
I_{\vartheta}\, N\!\left(d H_c + L_c H_c^2 + H_c m\right)
\right).
\]
For fixed $d$, $L_c$, and $H_c$, and moderate-to-large $N$, the $N^3$ Cholesky term dominates and the cost simplifies to $\mathcal{O}\big((I_{\vartheta}+I_{\phi}) N^3\big)$. Hence, if the block-coordinate procedure requires $T$ outer iterations to converge, the overall computational complexity of Algorithm~\ref{algo:param_estimation} for a given network architecture is
\(
\mathcal{O}\big(T (I_{\vartheta}+I_{\phi}) N^3\big).
\)

The memory requirements are similarly dominated by the $\mathcal{O}(N^2)$ storage of the dense covariance matrix and its Cholesky factor. In contrast, storing the parameters of the neural autoregressive flow requires
\(
\mathcal{O}\!\left(d H_c + L_c H_c^2 + H_c m\right),
\)
which is typically negligible relative to the covariance storage since, in practice, $N^2$ is much larger than $d H_c + L_c H_c^2 + H_c m$.

\section{Simulation experiments}\label{sec:2_dim}

\begin{figure}[t!]
    \centering
        \includegraphics[]{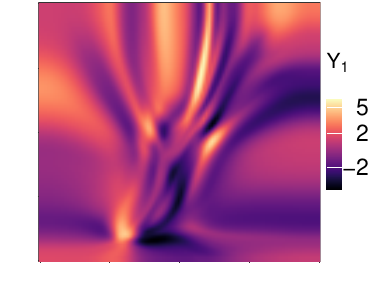}
      \includegraphics[]{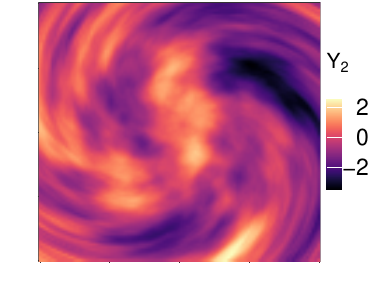}
    \caption{Simulated spatial fields $Y_1(\cdot)$ (left panel) and $Y_2(\cdot)$ (right panel). }
    \label{fig:True_process}
\end{figure}

In this section we evaluate our approach to spatial modeling by comparing it to other approaches through two simulation experiments. In the first experiment we generated data from a spatial input warped Gaussian process \citep[SIWGP, ][]{zammit2021deep}. The warping function in the SIWGP comprises axial warping units, which warp each spatial coordinate separately, nine radial basis function units, which locally warp the spatial domain, and a M{\"o}bius transformation, which acts globally on the domain $D = [-0.5, 0.5]^2$; for more details of these warping units see \citet{zammit2021deep}. In the second experiment we generated data from a Gaussian process under the following ``spiral'' warping function,
\begin{equation}\label{eq:third_warping}
   f(s_1, s_2) 
= 
r \bigl( 
\cos(\theta + a r + b r^{3}),\;
\sin(\theta + a r + b r^{3}) 
\bigr), 
\end{equation}
where $r = \sqrt{s_1^2 + s_2^2},$ and $\theta = \operatorname{atan2}(s_2, s_1),$ for $\bfs \in D$.  We denote the process associated with the first experiment as $Y_1(\cdot)$, and that with the second experiment as $Y_2(\cdot)$. We simulated both processes on a 101 $\times$ 101 gridding of $D$ (Figure \ref{fig:True_process}); we denote each of these grid points by $\bfs_{0,j}, j = 1,\ldots, n_p$, where $n_p = 101^2$, and collect these points in the set $D_0 \equiv \{\bfs_{0,1},\dots,\bfs_{0,n_p} \}$. For each simulation experiment, we simulated $N = 2000$ synthetic observations by sampling without replacement from the $n_p$ simulated process values, and adding measurement error with variance $\sigma^2_\epsilon  = 0.01$ (see \eqref{eq:obs}).

We compare our modeling approach to four other spatial process models. The first, which we refer to as GP$_\text{stat}$, is a mean zero stationary Gaussian process model with a Mat{\'e}rn covariance function 
\begin{equation}
\widetilde{C}_{\boldsymbol{\phi}}(h) = 
\phi_1^2 \cdot \dfrac{2^{1 - \phi_3}}{\Gamma(\phi_3)} 
\left( \dfrac{h}{\phi_2} \right)^{\phi_3} 
\mathcal{K}_{\phi_3} \!\left( \dfrac{h}{\phi_2} \right),
\label{eq:matern_cov}
\end{equation}
where $h = \|\bfs_i - \bfs_j\|, \bfs_i, \bfs_j \in D$, and where $\mathcal{K}_{\phi_3}$ is the modified Bessel function of the second kind of order $\phi_3$. The second model we consider, which we refer to as GP$_{\text{nonstat}}$, has a nonstationary Mat\'ern covariance function \citep{paciorek2006spatial}, where the parameters are spatially varying. Specifically, we model the spatially varying parameters as an inverse distance weighting, done using a squared exponential kernel, of parameter values at a few spatial nodes; details of our construction are given in Appendix \ref{sec:appendix_A} (see also \cite{li2019efficient} and \cite{nag2024_NS_Matern} for a similar model construction). In these simulation studies we use two nodes: $(-0.25, -0.25)'$ and $(0.25,0.25)'$. We also compare to the SIWGP process model, which we refer to as \( \text{GP}_{\text{deepspat}} \), and which represents the true process model for $Y_1(\cdot)$. For both $Y_1(\cdot)$ and $Y_2(\cdot)$ we use the same architecture for \( \text{GP}_{\text{deepspat}} \), which comprises two axial warping units (one for each spatial coordinate), nine radial basis functions, and a M{\"o}bius transformation. Lastly, we compare our modeling approach with DeepKriging \citep{nag2024bivariatedeepkriginglargescalespatial}, a spatially dependent deep neural network architecture suitable for modeling nonstationary spatial processes.

Our approach, GP$_\text{NAF}$, models the process as a mean zero Gaussian process on a warped domain with Mat\'ern covariance function given by \eqref{eq:matern_cov}, for $h = \|\bfs_i - \bfs_j\|, \bfs_i, \bfs_j \in \mathcal{D}$. We construct the normalizing flow $\mathbf{T}(\cdot)$ using two DDSF layers for each spatial coordinate, linked through composition. Each DDSF layer has the form given in \eqref{eq:ddsf}, and for each layer we set \( L = 5 \) sublayers. In each DDSF we set  \( \tilde{M}_l = M_{l} = 16,\) for \(l = 1, \ldots, 4\), and $\tilde{M}_5 = 16$. For the conditioner network we set $L_c = 5$ and $H_c = 100$. The parameter network \( \boldsymbol{\gamma}(\cdot;\boldsymbol{\vartheta}) \), responsible for outputting the parameters of the flow, is implemented as a feedforward neural network consisting of five fully connected layers. Each of these layers follows \eqref{eq:1layergamma} and uses $100$ hidden units.

We assess the predictive performance of each model using diagnostics evaluated at the locations in $D_0$: the locations at which the processes were simulated. We compute these diagnostics by comparing the probabilistic predictions from the five models to the true process values at those locations.  The diagnostics we consider are the mean squared prediction error (MSPE), the 95\% prediction interval coverage probability (PICP), and the mean prediction interval width (MPIW). These are defined as
\begin{align*}
    \text{MSPE} &=  {\dfrac{1}{n_p}\sum_{j=1}^{n_p} (Y(\bfs_{0,j}) - \hat{Y}(\bfs_{0,j}))^2},\\
     \text{PICP} &= \dfrac{1}{n_p}\sum_{j=1}^{n_p} \mathbbm{1}\{{Y(\bfs_{0,j})\in [L(\bfs_{0,j}),U(\bfs_{0,j})]}\}, \ \\
     \text{MPIW} &= \dfrac{1}{n_p}\sum_{j=1}^{n_p} [U(\bfs_{0,j}) - L(\bfs_{0,j})],
   \end{align*}
where $L(\bfs)$ and $U(\bfs)$ are the lower and upper prediction bounds of the 95\% prediction interval of $Y(\bfs)$. Computations involving GP$_{\text{stat}}$ and GP$_{\text{nonstat}}$ were done on a high-end desktop computer with an Intel\textsuperscript{\textregistered} Core\textsuperscript{TM} i9-14900HX CPU and 128\,GB of RAM. Computations involving GP$_{\text{NAF}}$ and GP$_{\text{deepspat}}$ made additional use of an NVIDIA\textsuperscript{\textregistered} GeForce RTX~4090 GPU.

\begin{table}[t!]
    \centering
    \caption{Performance comparison of different comparing models on the two simulated datasets. The evaluation metrics are MSPE, PICP, and MPIW and total training time (in mins). Lower MSPE and MPIW values indicate better performance, while PICP close to 0.95 reflects a well-calibrated prediction interval.}
    \vspace{0.5em}
    \begin{tabular}{|l l c c c c|}
    \hline
        Process & Model & MSPE & PICP & MPIW & Time (in mins)\\
        \hline
       \hline
        \multirow{3}{*}{$Y_1(\bfs)$} 
            & $\text{GP}_{\text{NAF}}$   & 0.027 & 0.95 & 0.42 & 37\\
            & $\text{GP}_{\text{deepspat}}$   & {0.019} & {0.95} & 0.39 & 45\\
            & $\text{GP}_{\text{nonstat}}$ & 0.097 & 0.88 & 1.48 & 27\\
            & DeepKriging     & 0.093 & 0.88 & 1.13 & 8\\
            & $\text{GP}_{\text{stat}}$     & 0.130 & 0.89 & 1.29 & 18\\    
        \hline
        \hline
        \multirow{3}{*}{$Y_2(\bfs)$} 
            & $\text{GP}_{\text{NAF}}$   & 0.001 & 0.95 & 0.13 & 31\\
            & $\text{GP}_{\text{deepspat}}$   & 0.003 & 0.93 & 0.18 & 39\\
            & $\text{GP}_{\text{nonstat}}$ & 0.003 & 0.82 & 0.11 & 23\\
            & DeepKriging     & 0.003 & 0.99 & 0.29 & 7\\
            & $\text{GP}_{\text{stat}}$     & 0.003 & 0.98 & 0.34 & 14\\

        \hline
    \end{tabular}
    \label{tab:1}
\end{table}

Table~\ref{tab:1} shows the results of the five competing models—\( \text{GP}_{\text{stat}} \), \( \text{GP}_{\text{nonstat}} \), \( \text{GP}_{\text{deepspat}} \), \( \text{GP}_{\text{NAF}} \), and DeepKriging, fitted to data from the two simulated spatial processes, \( Y_1(\cdot) \) and \( Y_2(\cdot) \). Our proposed model, \( \text{GP}_{\text{NAF}} \), achieves MSPE values that closely match those of the true model, \(\text{GP}_{\text{deepspat}}\), in the first scenario, while also providing comparably tight and well-calibrated prediction intervals. In contrast, \( \text{GP}_{\text{nonstat}} \), \( \text{GP}_{\text{stat}} \), and DeepKriging exhibit substantially higher MSPE and MPIW values, reflecting reduced predictive accuracy and wider uncertainty bands. In the second experiment, \( \text{GP}_{\text{NAF}} \) outperforms all competing methods by a considerable margin in MSPE, and is also the method that gives the most calibrated prediction intervals. These findings clearly show how \( \text{GP}_{\text{NAF}} \) is able to capture a wide range of spatial structures, without the need for prior physical knowledge or tailored architecture design.

\begin{figure}[t!]
    \centering
    \begin{tabular}{ccccc}  
        \adjustbox{valign=m,vspace=1pt}{\includegraphics[]{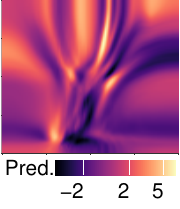}} &  
      \adjustbox{valign=m,vspace=1pt}{\includegraphics[]{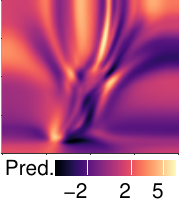}} & 
 \adjustbox{valign=m,vspace=1pt}{ \includegraphics[]{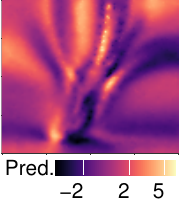}} &
 \adjustbox{valign=m,vspace=1pt}{\includegraphics[]{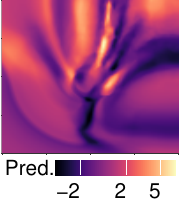}} & 
        \adjustbox{valign=m,vspace=1pt}{\includegraphics[]{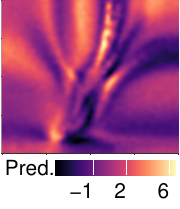}}
        \\
        \adjustbox{valign=m,vspace=1pt}{\includegraphics[]{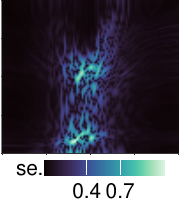}} & 
        \adjustbox{valign=m,vspace=1pt}{\includegraphics[]{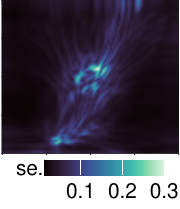}} & 
        \adjustbox{valign=m,vspace=1pt}{\includegraphics[]{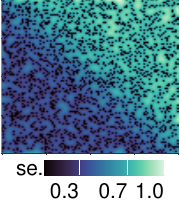}} & 
        \adjustbox{valign=m,vspace=1pt}{\includegraphics[]{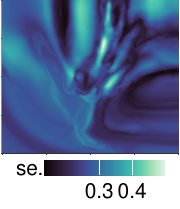}} &
        \adjustbox{valign=m,vspace=1pt}{\includegraphics[]{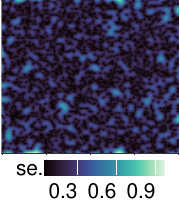}}
        \\       
    \end{tabular}
    \caption{Predicted spatial fields (top row) and their corresponding standard errors (bottom row) for the models (left to right) $\text{GP}_{\text{NAF}}$, $\text{GP}_{\text{deepspat}}$, $\text{GP}_{\text{nonstat}}$, DeepKriging, and $\text{GP}_{\text{stat}}$ used to model the process $Y_1(\cdot)$. }
    \label{fig:Y2}
\end{figure}

\begin{figure}[htbp]
    \centering
    \begin{tabular}{ccccc}  
        \adjustbox{valign=m,vspace=1pt}{\includegraphics[]{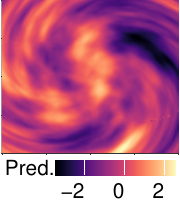}} &  
      \adjustbox{valign=m,vspace=1pt}{\includegraphics[]{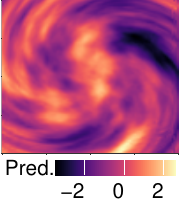}} & 
 \adjustbox{valign=m,vspace=1pt}{ \includegraphics[]{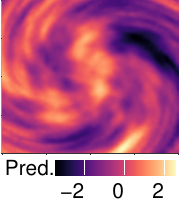}} & 
 \adjustbox{valign=m,vspace=1pt}{\includegraphics[]{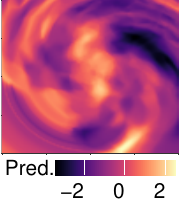}} &
        \adjustbox{valign=m,vspace=1pt}{\includegraphics[]{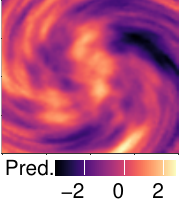}}
        \\
        \adjustbox{valign=m,vspace=1pt}{\includegraphics[]{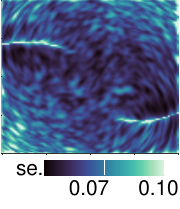}} & 
        \adjustbox{valign=m,vspace=1pt}{\includegraphics[]{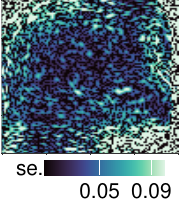}} & 
        \adjustbox{valign=m,vspace=1pt}{\includegraphics[]{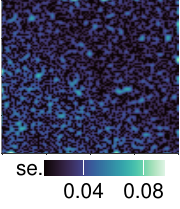}} & 
        \adjustbox{valign=m,vspace=1pt}{\includegraphics[]{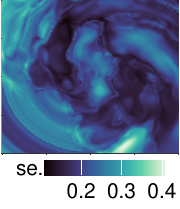}}&
        \adjustbox{valign=m,vspace=1pt}{\includegraphics[]{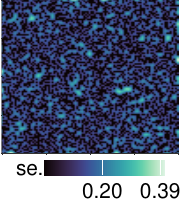}}
         \\       
    \end{tabular}
    \caption{Same as Figure~\ref{fig:Y2}, but for the process $Y_2(\cdot)$.}
    \label{fig:Y3}
\end{figure}

Figures~\ref{fig:Y2} and ~\ref{fig:Y3} show predictions from GP$_{\text{NAF}}$ and those from the other models. The standard error plots for GP\textsubscript{stat} and GP\textsubscript{nonstat} show patterning corresponding to observation locations, where predictive uncertainty reduces to nearly zero. In contrast, this patterning is not evident in the GP$_{\text{deepspat}}$, GP$_{\text{NAF}}$ and DeepKriging standard error plots, where uncertainty is instead dominated by the spatially varying properties of the process. These plots reflect more generally the ability of  warping-based methods to capture spatial distortions, such as stretching and compression, across the 2D domain in a manner consistent with the true generative process.

Figure~\ref{fig:optimization_plot} in Appendix~\ref{sec:appendix_B} presents an empirical visualization of the optimization process for Algorithm~\ref{algo:param_estimation} under the first simulation scenario, illustrating the convergence behavior of the proposed method.

\section{Real data application} \label{sec:real_data}

Accurate and spatially continuous maps of ocean temperature are an essential component of climate modeling, and are important for studying several Earth system processes \citep{stammer2000ocean, kirtman2012impact}. Ocean temperature plays a crucial role in regulating global climate by influencing heat distribution, atmospheric circulation, and biogeochemical processes \citep{schmitt2008salinity}. The Argo program, consisting of a global array of autonomous profiling floats, provides extensive subsurface ocean temperature and salinity observations \citep{roemmich2009argo}. However, despite the broad coverage and depth-resolving capabilities of the Argo profiling float network, observational gaps persist due to logistical constraints and uneven float distribution. These data gaps can hinder our ability to model and understand heterogeneous thermal variations, especially in dynamically active regions or at greater depths \citep{stammer2021well}.

To overcome these challenges, significant efforts have been made to construct high-resolution interpolated temperature fields from Argo observations, which are then used as boundary conditions, validation sources, or assimilation inputs in climate models \citep{lorenc1986analysis}. In particular, statistical methods that are able to quantify spatial uncertainty and naturally incorporate the spatial covariance structure of oceanographic processes are being increasingly used. For instance, the EN4 dataset, developed by the U.K. Met Office \citep{good2013}, employs optimal interpolation techniques grounded in statistical estimation theory to produce monthly gridded fields. Similarly, JAMSTEC's Monthly Objective Analysis using Argo (MOAA) \citep{hosoda2008} applies objective analysis schemes to generate temperature reconstructions, incorporating climatological baselines. 

\subsection{Exploratory data analysis}\label{sec:exp_data_analysis}

\begin{figure}[t!]
    \centering
        \includegraphics[]{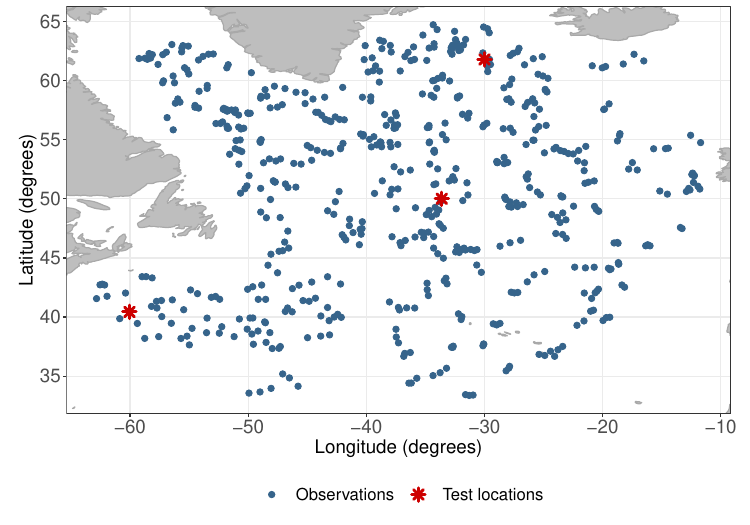} 
    \caption{Map of the Argo profiling network consisting of 604 locations in the Atlantic Ocean in January 2020, selected within a 2000 km radius centered at longitude $40^\circ$W and latitude $50^\circ$N. The three red markers indicate the test locations used for evaluation, located at $(29.993^\circ \text{W},\,61.802^\circ \text{N})$, $(60.052^\circ \text{W},\,40.430^\circ \text{N})$, and $(33.610^\circ \text{W},\,50.002^\circ \text{N})$.\vspace{0.2in}}
    \label{fig:excluded_locs_argo}
\end{figure}

In this study, we use the \texttt{argoFloats} R package \citep{argofloats} to analyze Argo profiling float data using Gaussian process (GP) regression. Our focus is on a region in the Atlantic Ocean with a radius of approximately 2000 km, centered at $(40^\circ\mathrm{W},\,50^\circ\mathrm{N})$, which contains 604 Argo floats, as shown in Figure~\ref{fig:excluded_locs_argo}, with data averaged over the time period of January 2020. This region was selected because of its pronounced nonstationary behavior (see Section \ref{sec:inferece_argo}), which provides a suitable setting for evaluating the ability of the proposed method to capture spatially varying dependence structures. 
We model ocean temperature while using pressure (measured in decibars, dbar) as a proxy for depth. For each spatial location we consider the temperature at ten pressure strata, obtained by discretizing the pressure interval for that location into 10 equidistant intervals and computing the average temperature within each stratum. This leads to a dataset consisting of 6040 observations. Prior to model fitting, the spatial coordinates are standardized to the domain $D \equiv [0,1]^3$ using a min--max transformation.

To remove large-scale deterministic variation, we first estimate and subtract a linear trend in temperature as a function of longitude, latitude, and pressure. Specifically, for $i = 1, \dots, N$, where $N$ is the total number of 3-dimensional locations, letting $Z_i$ denote the observed temperature at spatial location $\mathbf{s}_i \equiv (\text{lon}_i,\text{lat}_i,\text{pres}_i)'$, we fit the regression model
\begin{equation}
Z_i = \beta_0 + \beta_1 \, \text{lon}_i + \beta_2 \, \text{lat}_i + \beta_3 \, \text{pres}_i + \eta(\mathbf{s}_i) + \epsilon_i, \quad i = 1, \dots, N,
\end{equation}
where $\eta(\mathbf{s}_i)$ represents the residual spatial process. The residuals obtained from this model are then standardized and treated as the response for subsequent spatial modeling. Both the proposed method and all competing approaches are fitted to these standardized residuals.

To train the GP models on the standardized residual spatial process of the \texttt{argoFloats} dataset, we first exclude three spatial locations, across all pressure levels, from the dataset for location-specific testing. These three test sites are indicated by red star markers in Figure~\ref{fig:excluded_locs_argo}. From the remaining observations, we then randomly select 5409 three-dimensional locations for training and reserve 601 for validation when comparing the proposed method with competing models.

\subsection{Training and validation}\label{sec:inference_argo}
As in the simulation experiment, we compare our proposed model GP$_{\text{NAF}}$ to other competing models. However, we exclude GP$_{\text{deepspat}}$, since this cannot handle 3-dimensional locations. For GP$_{\text{NAF}}$ we use the same architecture for $\mathbf{T}(\cdot)$ as in the simulation studies. For GP$_{\text{nonstat}}$ we increase the number of nodes to three, with node points at locations  $(0.25,0.25,0.25)'$, $(0.25,0.75,0.25)'$ and $(0.75,0.75,0.75)'$ in $D$. Fitting GP$_{\text{NAF}}$ required $67$ minutes, mostly using the GPU, while fitting GP$_{\text{nonstat}}$ and GP$_{\text{stat}}$ required $54$ minutes and $51$ minutes with the CPU, respectively. Prediction diagnostics computed at the 601 validation locations are shown in  Table \ref{tab:2}. These results confirm that the flexibility of GP$_{\text{NAF}}$ is needed to get reliable probabilistic predictions of ocean temperature from Argo data.

\begin{table}[!t]
    \centering
    \caption{Performance comparison of four comparing models on the 3D Argo dataset based on 601 validation locations. Evaluation metrics include MSPE, PICP, and MPIW but computed with respect to the data $Z$ instead of the process $Y$. Lower MSPE and MPIW values indicate better predictive accuracy and tighter uncertainty bounds, while a PICP close to 0.95 suggests well-calibrated uncertainty. The compute time required for fitting the models is also shown.}
    \vspace{0.5em}
    \begin{tabular}{|l c c c c|}
        \hline
        Model & MSPE & PICP & MPIW & Time (in mins)\\
        \hline
             $\text{GP}_{\text{NAF}}$   & {0.17} & {0.95} & {0.37} & 67\\
             $\text{GP}_{\text{nonstat}}$ & 0.39 & 0.98 & 2.36 & 54\\
             DeepKriging     & 0.47 & 0.91 & 2.19 & 13\\
             $\text{GP}_{\text{stat}}$     & 0.51 & 0.95 & 2.22 & 51\\
        \hline
    \end{tabular}
    \label{tab:2}
\end{table}

\begin{figure}[t!]
    \centering
        \begin{tabular}{ccc}
            \includegraphics[]{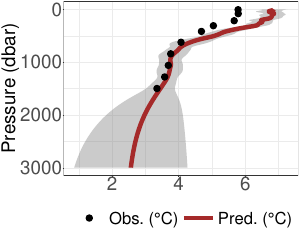} &
            \includegraphics[]{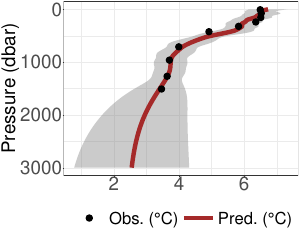} &
            \includegraphics[]{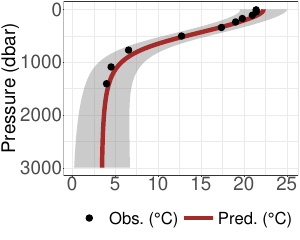}
        \end{tabular}
    \caption{Predictions and corresponding prediction intervals ($\pm$ 1.96 standard errors) of data ($Z$)
 across the full range of pressure levels (in dbar) for the three test locations (from left to right): $(29.9930^\circ \text{W}, \, 61.8020^\circ \text{N})$, $(60.0520^\circ\text{W}, \, 40.4300^\circ\text{N})$, and $(33.6100^\circ\text{W},\,50.0020^\circ\text{N})$.}
    \label{fig:sin_loc_preds}
\end{figure}

\begin{figure}[t!]
    \centering
    \includegraphics[width=\textwidth]{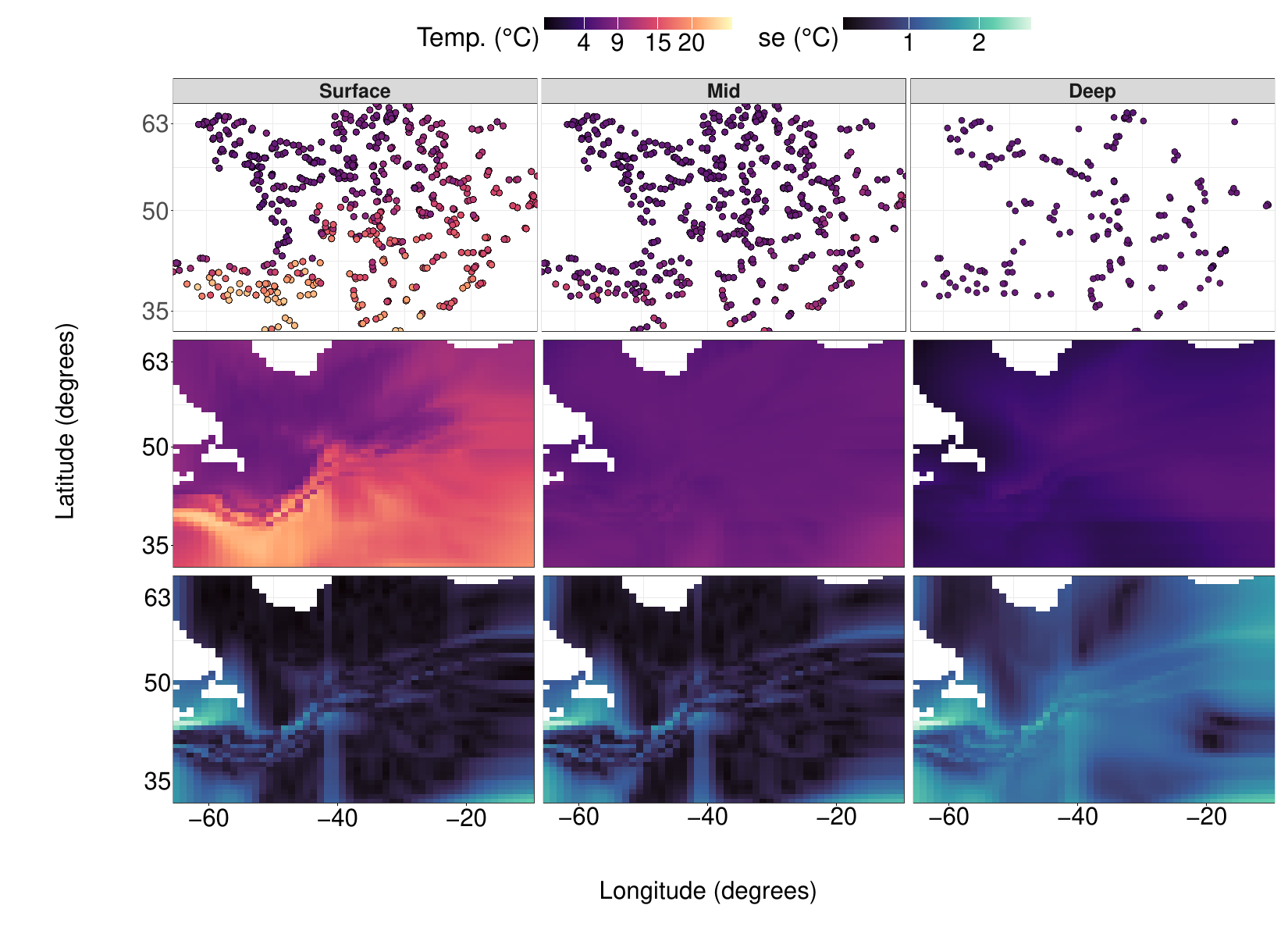}
    \caption{Top row: Observed \texttt{argoFloats} data points at three pressure levels (from left to right): 0 dbar, 1500 dbar, and 2700 dbar. Middle row: Spatial interpolation with GP$_{\text{NAF}}$ of ocean temperature. Bottom row: Corresponding standard errors. The horizontal and vertical axes represent longitude and latitude, respectively.}
    \label{fig:sing_pres_argo}
\end{figure}

Next, we take a close look at the predictions from $\text{GP}_{\text{NAF}}$ at the three test sites shown in Figure \ref{fig:excluded_locs_argo} that were also not used for model fitting. The predictions at these locations are presented in Figure~\ref{fig:sin_loc_preds}. 
At these sites we only have temperature data up to a pressure level of 1500 dbar; beyond this depth, the model extrapolates by leveraging information from surrounding spatial locations. Due to the sparsity of observations at deeper levels, the prediction intervals widen accordingly. All test data, except the shallow ones of the first test depth profile, are well-within the 95\% prediction intervals. The inaccurate predictions at the lower pressures of the first test profile are due to the depth profiles in close proximity to the test location all having substantially warmer shallow-water temperatures; see Figure \ref{fig:excluded_loc_plot} in Appendix \ref{sec:appendix_C}, where we plot the test data together with the observations from the nearest five depth profiles. Finally, Figure~\ref{fig:sing_pres_argo} shows spatial predictions at three pressure levels across the study region. These predictions clearly show the heterogeneity in ocean temperature by depth, and the need for a model that can capture the spatially-varying anisotropy  and the inherent complexity of this physical process. Code for reproducing the results in Sections \ref{sec:2_dim} and \ref{sec:real_data} is available from 
\url{https://github.com/pratiknag/Spatial_NormalizingFlows_Code.git}.

\subsection{Inference with GP$_{\text{NAF}}$}\label{sec:inferece_argo}

\begin{figure}[t!]
    \centering
    \begin{subfigure}[t]{0.48\textwidth}
        \centering
        \includegraphics[width=\textwidth]{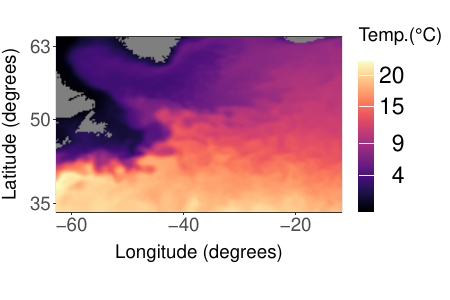}
        \caption{}
        \label{fig:sst}
    \end{subfigure}
    \hfill
    \begin{subfigure}[t]{0.48\textwidth}
        \centering
        \includegraphics[width=\textwidth]{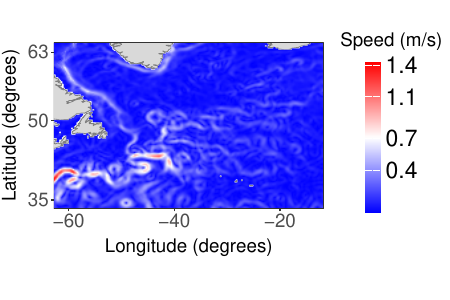}
        \caption{}
        \label{fig:currents_only}
    \end{subfigure}
    \caption{(a) Sea surface temperature (SST) field from the  PODAAC data product over the study region for January 2020. (b) Corresponding surface ocean current speed.}
    \label{fig:surface_currents}
\end{figure}

Here, we compare predictions from the proposed GP$_{\text{NAF}}$ model against sea surface temperature (SST) fields obtained from the PODAAC data product \citep{mur_sst_2015}. Figure~\ref{fig:sst} illustrates the SST distribution over the selected region for January 2020, where we can see ample evidence of second-order nonstationarity, with fast changing temperatures in the South-West, and slowly varying temperatures in the North-East. A visual comparison between the center-left panel of Figure~\ref{fig:sing_pres_argo} and Figure \ref{fig:sst} indicates that the GP$_{\text{NAF}}$ reconstruction recovers the large-scale spatial patterns in PODAAC. While there are differences in the predicted fine-scale variability, there is a marked correspondence between the areas of fast flow in Figure~\ref{fig:currents_only} and the regions of high variability in the center-left panel of Figure~\ref{fig:sing_pres_argo}.

\begin{figure}[t!]
    \centering
    \includegraphics[]{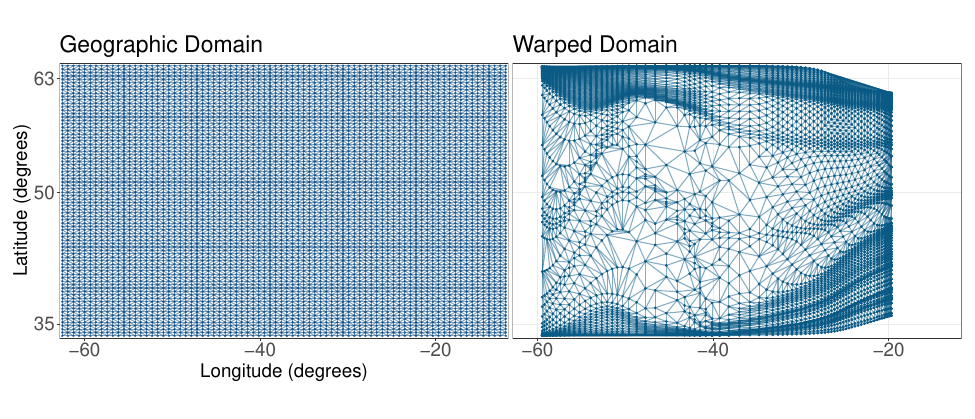}
    \caption{Learned spatial warping of the ocean surface. The left panel shows points in the geographic domain, while the right panel shows the corresponding points in the warped domain, as inferred by GP$_{\text{NAF}}$.}
    \label{fig:surface_warping}
\end{figure}

The learned warping function, shown in Figure~\ref{fig:surface_warping}, is reflective of the surface current velocities in PODAAC.  In particular, the region centered approximately at $35^\circ\text{W}$ longitude and $40^\circ\text{N}$ latitude exhibits relatively strong surface currents, whereas the flow weakens as we move eastward toward $30^\circ\text{W}$. 
Correspondingly, regions with stronger currents are stretched in the warped domain, while regions with weaker currents are compressed. Intuitively, high-velocity regions are associated with rapid spatial changes in temperature, resulting in smaller correlation length scales. Conversely, low-velocity regions exhibit smoother temperature variations and hence larger correlation length scales. The warping function effectively homogenizes these varying dependence structures, enabling the use of a stationary covariance model in the transformed space.

\begin{figure}[t!]
    \centering
    \begin{minipage}{0.48\textwidth}
        \centering
        \includegraphics[width=\textwidth]{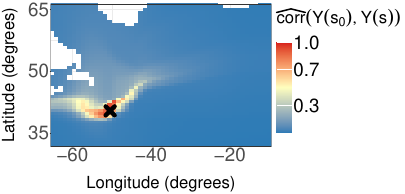}
    \end{minipage}
    \hfill
    \begin{minipage}{0.48\textwidth}
        \centering
        \includegraphics[width=\textwidth]{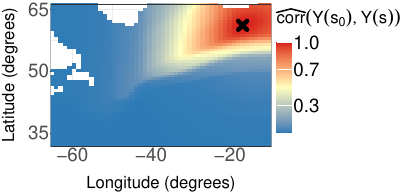}
    \end{minipage}
    \caption{Spatial correlation structure induced by GP$_{\text{NAF}}$ at two representative locations marked by a black cross. The left panel shows the inferred correlation with respect to $(56^\circ\text{W}, 40^\circ\text{N})$, while the right panel shows the inferred correlation with respect to $(20^\circ\text{W}, 63^\circ\text{N})$.}
    \label{fig:correlation_argo}
\end{figure}

The spatial variability in current magnitude induces pronounced nonstationarity in the SST field. Figure~\ref{fig:correlation_argo} shows the spatial correlation between selected reference points and all other locations in the domain as implied by the GP$_{\text{NAF}}$ model. The location at $(35^\circ\text{W}, 40^\circ\text{N})$, characterized by stronger currents, exhibits a rapid correlation decay, which is indicative of short-range dependence. In contrast, the location at $(20^\circ\text{W}, 63^\circ\text{N})$, where currents are weaker, displays a slower decay and hence longer-range spatial dependence. These results highlight the ability of GP$_{\text{NAF}}$ to capture spatially varying dependence structures.

\section{Discussion and future work} \label{sec:discussion}

GP$_{\text{NAF}}$ is a novel statistical spatial modeling framework specifically designed to model spatial processes characterized by complex, nonstationary, and anisotropic covariance structures. Unlike traditional geostatistical methods, which often assume stationarity or rely on rigid parametric forms, GP$_{\text{NAF}}$ leverages the expressive power of neural networks to learn flexible, data-driven spatial transformations. Central to its architecture is the use of normalizing flows, a sequence of invertible, differentiable mappings, for warping the spatial domain. This framework, facilitated through NAFs, inherently enforces injectivity, thereby ensuring smooth, one-to-one mappings of the input space. Such a constraint prevents pathological behaviors like space-folding, which can compromise the interpretability and validity of spatial predictions. The warping also provides interpretable insights into spatial deformations through local stretches, compressions, and directional anisotropies.
The NAFs can be contrasted with vanilla feedforward networks which, while expressive, do not naturally enforce invertibility and thus lack a clear geometric interpretation of their outputs.

GP$_{\text{NAF}}$ extends the GP$_{\text{deepspat}}$ modeling framework of \cite{zammit2021deep}, which was designed for 1D and 2D spatial processes and required choosing distinct warping units: GP$_{\text{NAF}}$ can be used to model processes in higher-dimensional spatial domains with relative ease. This flexibility opens the door to a wide range of applications spanning the geosciences and environmental sciences. 
From a computational standpoint, the implementation of GP$_{\text{NAF}}$ benefits from modern deep learning ecosystems such as PyTorch, which facilitate efficient gradient-based training and inference via automatic differentiation and GPU acceleration. These capabilities are crucial for scaling the method to large spatial datasets, which are prevalent in modern geostatistical applications.

A natural extension of the proposed framework is the incorporation of physics-guided constraints into the warping function. While our current approach is fully data-driven and does not rely on prior physical knowledge, it can be augmented to encode domain-specific structure through the loss function. For example, monotonicity constraints (e.g., with respect to depth) could be enforced by appropriately constraining the process on the warped domain \citep{david_dunson_monotone_GP2014}. More general physical relationships such as advection-diffusion dynamics could be incorporated via soft or hard constraints inspired by physics-informed neural networks (PINNs). Recent work connecting PINNs with statistical modeling provides a principled pathway for integrating such constraints within probabilistic frameworks \citep{Wikle_2026}. Embedding these constraints into the normalizing flow architecture would allow the learned warping to respect known physical laws while retaining flexibility, although doing so will likely introduce additional computational and architectural challenges. Finally, we note that our proposed model and these extensions are not limited to NAFs, and alternative normalizing flow models could be used within our framework.

Future work will consider a spatio-temporal extension of GP$_{\text{NAF}}$. Incorporating temporal dynamics into the flow architecture, for instance through temporal embeddings, would enable the model to capture evolving spatial patterns and dependencies over time. Such an extension would broaden the applicability of GP$_{\text{NAF}}$ to settings such as climate modeling, environmental monitoring, and real-time forecasting, where both spatial and temporal structures play a critical role.

\section*{Acknowledgements}

This material is based upon work supported by the Air Force Office of Scientific Research under award number FA2386-23-1-4100 (P.N. and A.Z.-M.). P.N. and Y.S.~acknowledge the support of King Abdullah University of Science and Technology (KAUST).

\appendix
\section{Nonstationary Mat\'ern covariance and kernel smoothing}\label{sec:appendix_A}

Consider a univariate Gaussian random field \( \{Y(\bfs): \bfs \in D \subset \mathbb{R}^d\} \), where \( Y(\cdot) \) is a zero-mean Gaussian process with covariance function \( C^{\text{NS}}(\cdot, \cdot) \). For GP$_{\text{nonstat}}$, we employ the nonstationary Mat\'ern covariance function \citep{paciorek2006spatial}
\begin{equation*}
C^\text{NS}(\bfs_i, \bfs_j) = \dfrac{\sigma(\bfs_i)\sigma(\bfs_j)|\boldsymbol{\Sigma}(\bfs_i)|^{1/4}|\boldsymbol{\Sigma}(\bfs_j)|^{1/4}}{\Gamma(\bar{\nu})2^{\bar{\nu}-1}} \left|\dfrac{\boldsymbol{\Sigma}(\bfs_i) + \boldsymbol{\Sigma}(\bfs_j)}{2}\right|^{-1/2} \left(2\sqrt{\bar{\nu}Q_{ij}}\right)^{\bar{\nu}} \mathcal{K}_{\bar{\nu}}(2\sqrt{\bar{\nu}Q_{ij}}),
\end{equation*}
where \( \bar{\nu} = \frac{\nu(\bfs_i) + \nu(\bfs_j)}{2} \), \( Q_{ij} \) is the Mahalanobis distance based on local anisotropy matrices \( \boldsymbol{\Sigma}(\bfs) \), and \( \mathcal{K}_{\bar{\nu}} \) is the modified Bessel function of the second kind and order $\bar{\nu}$. We express anisotropy via eigen decomposition of \( \boldsymbol{\Sigma}(\bfs) \) with orientation angle \( \alpha(\bfs) \) and eigenvalues \( \lambda_1(\bfs),\dots, \lambda_d(\bfs) \). All covariance parameters may vary spatially, defining a location-specific parameter vector \( \boldsymbol{\phi}(\bfs) = \left(\sigma(\bfs), \lambda_1(\bfs), \dots, \lambda_d(\bfs), \alpha(\bfs), \nu(\bfs)\right)' \), for $\bfs \in D$. To reduce the complexity of estimating these parameters across space, we employ a kernel smoothing approach: the parameters at any location \( \bfs_i \) are estimated via a weighted average of the parameter values at $K$ nodes \( \{\bfs^*_k\}_{k=1}^K \), where each \(\bfs^*_k \in D\), through
\[
\boldsymbol{\phi}(\bfs_i) = \sum_{k=1}^K W(\bfs_i, \bfs^*_k)\boldsymbol{\phi}(\bfs^*_k), \quad \text{with} \quad W(\bfs_i, \bfs^*_k) = \dfrac{\mathcal{C}(\bfs_i, \bfs^*_k)}{\sum_{k'=1}^K \mathcal{C}(\bfs_i, \bfs^*_{k'})},
\]
where \( \mathcal{C}(\bfs, \bfs^*_k) = \exp\left(-\|\bfs - \bfs^*_k\|^2 / (2h)\right) \) is a squared exponential kernel with bandwidth \( h \). The problem then reduces to estimating the parameters at the nodes, $\boldsymbol{\phi}(\bfs^*_k)$, for $k = 1,\dots, K$. For computational feasibility, we assume constant \( \alpha \), and constrain \(\lambda_1(\bfs) = \cdots = \lambda_d(\bfs) \equiv \lambda(\bfs), \) for  \(\bfs \in D\). Readers are referred to \cite{nag2024_NS_Matern} for a study involving this model construction.

\section{Numerical stability and optimization}\label{sec:appendix_B}

In this section, we discuss further the optimization of the NAF for the simulation scenario involving $Y_1(\cdot)$. Recall that Algorithm~\ref{algo:param_estimation} performs two separate optimization steps within each outer loop. Since gradient-based optimization for Gaussian processes is well studied in the literature, we focus here on the optimization of the NAF parameters.

\begin{figure}[t!]
    \centering
        \includegraphics[]{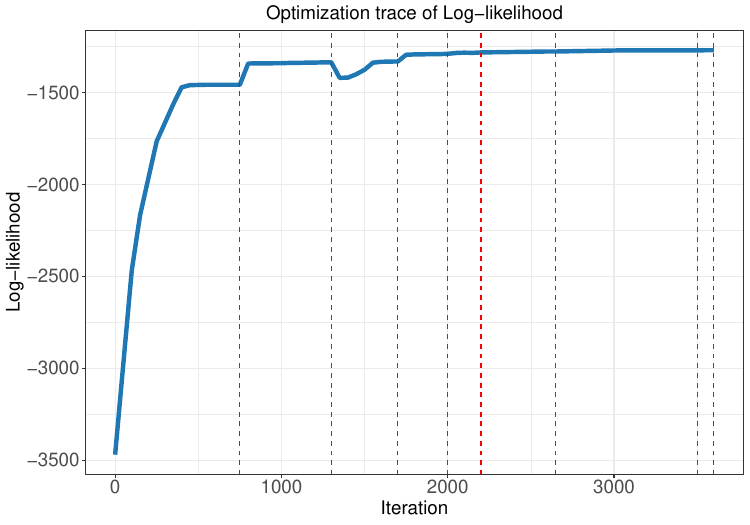} 
    \caption{The log-likelihood as a function of inner loop iteration number when training GP$_{\text{NAF}}$ in the first simulation experiment of Section~\ref{sec:2_dim}. 
The segments separated by the red dashed lines correspond to end-points of outer loop iterations in Algorithm~\ref{algo:param_estimation}. }
    \label{fig:optimization_plot}
\end{figure}

Figure~\ref{fig:optimization_plot} displays the log-likelihood as a function of the iterations, with vertical red dashed lines indicating the end of each outer loop. The figure shows that the the log-likelihood converges by the eighth outer iteration. Within each outer loop, the log-likelihood jumps  at the start; this is due to the optimization of the GP parameters which is done at the beginning of each outer loop iteration. While this should always lead to an increase in the log-likelihood, we see that this is not always the case, as seen in the third outer iteration, where the log-likelihood actually decreases. This transient dip is attributed to the use of stochastic gradients in the optimization of the NAF parameters, which may momentarily move the parameters away from a locally optimal direction. Such behavior is common in stochastic optimization, and it reflects the inherent trade-off between exploration and stability during training.
\section{Additional figures}\label{sec:appendix_C}
\begin{figure}[t!]
    \centering
        \includegraphics[]{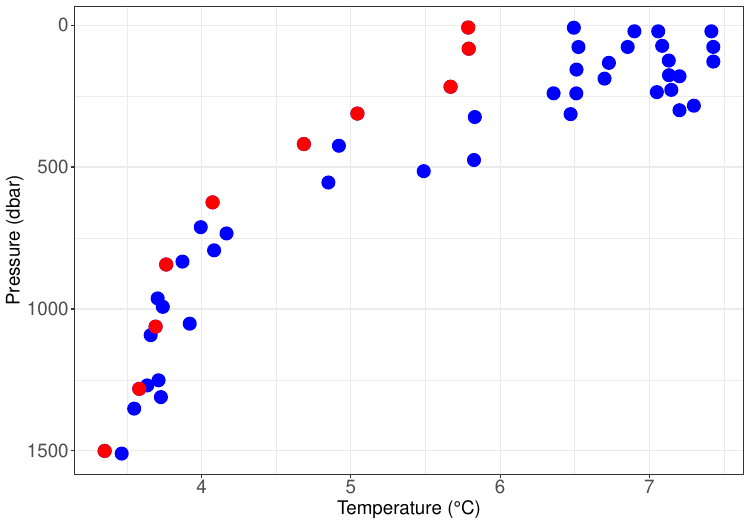} 
    \caption{Test data (red points) at $(29.993^\circ\text{W},\,61.802^\circ\text{N})$ and observations (blue points) from the five nearest depth profiles.}
    \label{fig:excluded_loc_plot}
\end{figure}

Figure \ref{fig:excluded_loc_plot} illustrates the test data at the location $(29.993^\circ\mathrm{W},\,61.802^\circ\mathrm{N})$, shown alongside observations from the five nearest depth profiles. This plot provides insight into the relatively poor prediction at this location as shown in Figure \ref{fig:sin_loc_preds}, left panel.

\clearpage
\bibliographystyle{apalike}

\bibliography{ref}

@article{Zammit_2021,
  title={{FRK}: An {R} package for spatial and spatio-temporal prediction with large datasets},
  author={Zammit-Mangion, Andrew and Cressie, Noel},
  journal={Journal of Statistical Software},
  volume={98},
  pages={1--48},
  year={2021}
}

@article{Wikle_2026,
author = {Christopher K. Wikle and Joshua North and Giri Gopalan and Myungsoo Yoo},
title = {A Statistician’s Overview of Physics-Informed Neural Networks for Spatio-Temporal Data},
journal = {Journal of the American Statistical Association},
volume = {121},
number = {554},
pages = {1708--1724},
year = {2026},
publisher = {Taylor \& Francis},
doi = {10.1080/01621459.2026.2625420},


URL = { 
    
        https://doi.org/10.1080/01621459.2026.2625420
    
    

},
eprint = { 
    
        https://doi.org/10.1080/01621459.2026.2625420
    
    

}

}

@article{rocha2018arctic,
  author  = {Rocha, A. V. and Blakely, B. and Jiang, Y. and Wright, K. S. and Curasi, S. R.},
  title   = {Is {A}rctic greening consistent with the ecology of tundra? {L}essons from an ecologically informed mass balance model},
  journal = {Environmental Research Letters},
  volume  = {13},
  number  = {12},
  pages   = {125007},
  year    = {2018},
  doi     = {10.1088/1748-9326/aaeb50}
}

@article{katzfuss2021general,
  author  = {Katzfuss, Matthias and Guinness, Joseph},
  title   = {A General Framework for {V}ecchia Approximations of {G}aussian Processes},
  journal = {Statistical Science},
  volume  = {36},
  number  = {1},
  pages   = {124--141},
  year    = {2021},
  doi     = {10.1214/19-STS755}
}

@article{guinness2018permutation,
  author  = {Guinness, Joseph},
  title   = {Permutation and Grouping Methods for Sharpening {G}aussian Process Approximations},
  journal = {Technometrics},
  volume  = {60},
  number  = {4},
  pages   = {415--429},
  year    = {2018},
  doi     = {10.1080/00401706.2018.1437476}
}

@misc{rocha2019modis,
  author       = {Rocha, A. V. and Wright, K. S.},
  title        = {{ABoVE: MODIS- and CCAN-Derived NDVI and Trends, North Slope of Alaska, 2000--2015}},
  year         = {2019},
  publisher    = {ORNL DAAC},
  address      = {Oak Ridge, Tennessee, USA},
  note          = {\url{https://doi.org/10.3334/ORNLDAAC/1666}}
}

@article{cressie2008fixed,
  title={Fixed rank kriging for very large spatial data sets},
  author={Cressie, Noel and Johannesson, Gardar},
  journal={Journal of the Royal Statistical Society B},
  volume={70},
  number={1},
  pages={209--226},
  year={2008},
  publisher={Oxford University Press}
}

@book{cressie1993statistics,
  title={Statistics for Spatial Data},
  author={Cressie, Noel},
  year={1993},
  publisher={John Wiley \& Sons, Hoboken, NJ, revised edition}
}

@article{stammer2000ocean,
  title={Ocean state estimation and prediction in support of oceanographic research},
  author={Stammer, Dethf and Chassignet, EP},
  journal={Oceanography},
  volume={13},
  number={2},
  pages={51--56},
  year={2000},
  publisher={The Oceanography Society}
}

@article{kirtman2012impact,
  author = {Kirtman, Ben P. and Bitz, Cecilia and Bryan, Frank and Collins, William and Dennis, John and Hearn, Nathan and Kinter, James L. and Loft, Richard and Rousset, Clement and Siqueira, Leo and Stan, Cristiana and Tomas, Robert and Vertenstein, Mariana},
  title = {Impact of ocean model resolution on {CCSM} climate simulations},
  journal = {Climate Dynamics},
  volume = {39},
  number = {6},
  pages = {1303--1328},
  year = {2012},
  doi = {10.1007/s00382-012-1500-3},
  url = {https://doi.org/10.1007/s00382-012-1500-3},
  issn = {1432-0894}
}

@article{schmitt2008salinity,
  title={Salinity and the global water cycle},
  author={Schmitt, RaymoNd W},
  journal={Oceanography},
  volume={21},
  number={1},
  pages={12--19},
  year={2008},
  publisher={JSTOR}
}

@article{lorenc1986analysis,
  title={Analysis methods for numerical weather prediction},
  author={Lorenc, Andrew C},
  journal={Quarterly Journal of the Royal Meteorological Society},
  volume={112},
  number={474},
  pages={1177--1194},
  year={1986},
  publisher={Wiley Online Library}
}

@article{stammer2021well,
  title={How well do we know ocean salinity and its changes?},
  author={Stammer, D and Martins, M Sena and K{\"o}hler, J and K{\"o}hl, A},
  journal={Progress in Oceanography},
  volume={190},
  pages={102478},
  year={2021},
  publisher={Elsevier}
}

@article{Vu_2025,
  title={deepspat: An {R} package for modeling nonstationary spatial and spatio-temporal {G}aussian and extremes data through deep deformations},
  author={Vu, Quan and Shao, Xuanjie and Huser, Rapha{\"e}l and Zammit-Mangion, Andrew},
  journal={arXiv preprint arXiv:2512.08137},
  year={2025}
}

@article{datta2016hierarchical,
  title={Hierarchical nearest-neighbor {G}aussian process models for large geostatistical datasets},
  author={Datta, Abhirup and Banerjee, Sudipto and Finley, Andrew O and Gelfand, Alan E},
  journal={Journal of the American Statistical Association},
  volume={111},
  number={514},
  pages={800--812},
  year={2016},
  publisher={Taylor \& Francis}
}

@article{lindgren2011explicit,
  title={An explicit link between {G}aussian fields and {Gaussian Markov} random fields: the stochastic partial differential equation approach},
  author={Lindgren, Finn and Rue, H{\aa}vard and Lindstr{\"o}m, Johan},
  journal={Journal of the Royal Statistical Society B},
  volume={73},
  number={4},
  pages={423--498},
  year={2011},
  publisher={Wiley Online Library}
}

@article{sampson1992nonparametric,
  title={Nonparametric estimation of nonstationary spatial covariance structure},
  author={Sampson, Paul D and Guttorp, Peter},
  journal={Journal of the American Statistical Association},
  volume={87},
  number={417},
  pages={108--119},
  year={1992},
  publisher={Taylor \& Francis}
}

@article{wang2024spatialdeepconvolutionalneural,
      title={Spatial Deep Convolutional Neural Networks}, 
      title = {Spatial deep convolutional neural networks},
journal = {Spatial Statistics},
volume = {66},
pages = {100883},
year = {2025},
issn = {2211-6753},
doi = {https://doi.org/10.1016/j.spasta.2025.100883},
url = {https://www.sciencedirect.com/science/article/pii/S2211675325000053},
author = {Qi Wang and Paul A. Parker and Robert Lund} 
}

@article{zammit2021deep,
  author = {Andrew Zammit-Mangion and Tin Lok James Ng and Quan Vu and Maurizio Filippone},
title = {Deep Compositional Spatial Models},
journal = {Journal of the American Statistical Association},
volume = {117},
number = {540},
pages = {1787--1808},
year = {2022},
publisher = {ASA Website},
doi = {10.1080/01621459.2021.1887741},


URL = { 
    
        https://doi.org/10.1080/01621459.2021.1887741
    
    

},
eprint = { 
    
        https://doi.org/10.1080/01621459.2021.1887741
    
    

}
}

@inproceedings{germain2015made,
  title={{MADE}: Masked autoencoder for distribution estimation},
  author={Germain, Mathieu and Gregor, Karol and Murray, Iain and Larochelle, Hugo},
  booktitle={International Conference on Machine Learning},
  pages={881--889},
  year={2015},
  organization={PMLR}
}

@article{dinh2015nice,
  title={{NICE}: Non-linear independent components estimation},
  author={Dinh, Laurent and Krueger, David and Bengio, Yoshua},
  journal={arXiv preprint arXiv:1410.8516},
  year={2014}
}

@inproceedings{huang2018neural,
  title={Neural autoregressive flows},
  author={Huang, Chin-Wei and Krueger, David and Lacoste, Alexandre and Courville, Aaron},
  booktitle={International Conference on Machine Learning},
  pages={2078--2087},
  year={2018},
  organization={PMLR}
}

@techreport{smith1996estimating,
  author = {Richard L. Smith},
  title = {Estimating Nonstationary Spatial Correlations},
  year = {1996},
  note = {Available from \url{https://rls.sites.oasis.unc.edu/postscript/rs/nonstationary.pdf}}
}

@inproceedings{perrin1999modelling,
  title={Modelling of non-stationary spatial structure using parametric radial basis deformations},
  author={Perrin, Olivier and Monestiez, Pascal},
  booktitle={geoENV II—Geostatistics for Environmental Applications: Proceedings of the Second European Conference on Geostatistics for Environmental Applications},
  editor={Jaime G{\'o}mez-Hern{\'a}ndez and A. O. Soares and Roland Froidevaux},
  pages={175--186},
  year={1999},
  organization={Springer Science and Business Media},
  address={Dordrecht, The Netherlands}
}

@article{schmidt2003bayesian,
  title={Bayesian inference for non-stationary spatial covariance structure via spatial deformations},
  author={Schmidt, Alexandra M and O'Hagan, Anthony},
  journal={Journal of the Royal Statistical Society B},
  volume={65},
  number={3},
  pages={743--758},
  year={2003},
  publisher={Oxford University Press}
}

@article{castro2013state,
  title={State space models with spatial deformation},
  author={Castro Morales, Fidel Ernesto and Gamerman, Dani and Paez, Marina Silva},
  journal={Environmental and Ecological Statistics},
  volume={20},
  pages={191--214},
  year={2013},
  publisher={Springer}
}

@book{banerjee2003hierarchical,
  title={Hierarchical Modeling and Analysis for Spatial Data},
  author={Banerjee, Sudipto and Carlin, Bradley P and Gelfand, Alan E},
  year={2003},
  publisher={Chapman and Hall/CRC, Boca Raton, FL}
}

@article{roemmich2009argo,
  title={The {A}rgo Program: Observing the global ocean with profiling floats},
  author={Roemmich, Dean and Johnson, Gregory C and Riser, Stephen and Davis, Russ and Gilson, John and Owens, W Brechner and Garzoli, Silvia L and Schmid, Claudia and Ignaszewski, Mark},
  journal={Oceanography},
  volume={22},
  number={2},
  pages={34--43},
  year={2009},
  publisher={JSTOR}
}

@article{ghulam_warping2021,
author = {Ghulam A. Qadir and Ying Sun and Sebastian Kurtek},
title = {Estimation of Spatial Deformation for Nonstationary Processes via Variogram Alignment},
journal = {Technometrics},
volume = {63},
number = {4},
pages = {548-561},
year = {2021},
publisher = {Taylor & Francis},
doi = {10.1080/00401706.2021.1883481},
}

@incollection{higdon2022non,
  booktitle={Bayesian Statistics 6: Proceedings of the Sixth Valencia International Meeting June 6-10, 1998},
  title={Non-stationary spatial modeling},
  editor={J M Bernardo and J O Berger and A P Dawid and A F M Smith},
  author={Higdon, Dave and Swall, Jenise and Kern, John},
  pages={761-768},
  publisher={Oxford University Press},
  address={Oxford, UK},
  year={1999}
}

@article{stein2005space,
  title={Space--time covariance functions},
  author={Stein, Michael L},
  journal={Journal of the American Statistical Association},
  volume={100},
  number={469},
  pages={310--321},
  year={2005},
  publisher={Taylor \& Francis}
}

@article{chen2020deepkriging,
  author = {Chen, Wanfang and Li, Yuxiao and Reich, Brian J and Sun, Ying},
  title = {{DeepKriging}: Spatially Dependent Deep Neural Networks for Spatial Prediction},
  journal = {Statistica Sinica},
  volume = {34},
  number = {1},
  pages = {291--311},
  year = {2024}
 }

@article{nag2023spatio,
  title={Spatio-temporal {DeepKriging} for interpolation and probabilistic forecasting},
  author={Nag, Pratik and Sun, Ying and Reich, Brian J},
  journal={Spatial Statistics},
  volume={57},
  pages={100773},
  year={2023},
  publisher={Elsevier}
}

@article{nag2024bivariatedeepkriginglargescalespatial,
      author = {Pratik Nag and Ying Sun and Brian J Reich},
title = {Bivariate {DeepKriging} for Large-Scale Spatial Interpolation of Wind Fields},
journal = {Technometrics},
volume = {67},
number = {3},
pages = {397--408},
year = {2025},
publisher = {Taylor \& Francis},
doi = {10.1080/00401706.2025.2453197},
URL = { 
    
        https://doi.org/10.1080/00401706.2025.2453197
},
eprint = { 
    
        https://doi.org/10.1080/00401706.2025.2453197
}
}

@misc{mur_sst_2015,
  author       = {{JPL MUR MEaSUREs Project}},
  title        = {{GHRSST Level 4 MUR Global Foundation Sea Surface Temperature Analysis (v4.1)}},
  year         = {2015},
  version      = {4.1},
  publisher    = {PO.DAAC},
  address      = {Pasadena, CA, USA},
  doi          = {10.5067/GHGMR-4FJ04},
  url          = {https://doi.org/10.5067/GHGMR-4FJ04},
  note         = {Dataset accessed [2026-02-13]}
}

@article{david_dunson_monotone_GP2014,
 ISSN = {00063444},
 URL = {http://www.jstor.org/stable/43305615},
  author = {Lizhen Lin and David B. Dunson},
 journal = {Biometrika},
 number = {2},
 pages = {303--317},
 publisher = {[Oxford University Press, Biometrika Trust]},
 title = {Bayesian monotone regression using {G}aussian process projection},
 urldate = {2026-03-24},
 volume = {101},
 year = {2014}
}

@inproceedings{papamakarios2017masked,
  author = {Papamakarios, George and Pavlakou, Theo and Murray, Iain},
 booktitle = {Advances in Neural Information Processing Systems},
 editor = {I. Guyon and U. Von Luxburg and S. Bengio and H. Wallach and R. Fergus and S. Vishwanathan and R. Garnett},
 pages = {2339-2348},
 publisher = {Curran Associates, Inc.},
 address = {Red Hook, NY},
 title = {Masked Autoregressive Flow for Density Estimation},
 url = {https://proceedings.neurips.cc/paper_files/paper/2017/file/6c1da886822c67822bcf3679d04369fa-Paper.pdf},
 volume = {30},
 year = {2017}
}

@article{kobyzev2020normalizing,
  title={Normalizing flows: An introduction and review of current methods},
  author={Kobyzev, Ivan and Prince, Simon JD and Brubaker, Marcus A},
  journal={IEEE Transactions on Pattern Analysis and Machine Intelligence},
  volume={43},
  number={11},
  pages={3964--3979},
  year={2020},
  publisher={IEEE}
}

@article{risser2016review,
      title={Review: Nonstationary Spatial Modeling, with Emphasis on Process Convolution and Covariate-Driven Approaches}, 
      author={Mark D. Risser},
      year={2016},
      journal={arXiv:1610.02447}
}

@article{paciorek2006spatial,
  title={Spatial modelling using a new class of nonstationary covariance functions},
  author={Paciorek, Christopher J and Schervish, Mark J},
  journal={Environmetrics},
  volume={17},
  number={5},
  pages={483--506},
  year={2006},
  publisher={Wiley Online Library}
}

@article{li2019efficient,
  title={Efficient estimation of nonstationary spatial covariance functions with application to high-resolution climate model emulation},
  author={Li, Yuxiao and Sun, Ying},
  journal={Statistica Sinica},
  volume={29},
  number={3},
  pages={1209--1231},
  year={2019},
  publisher={JSTOR}
}

@article{nag2024_NS_Matern,
author = {Pratik Nag and Yiping Hong and Sameh Abdulah and Ghulam A. Qadir and Marc G. Genton and Ying Sun},
title = {Efficient Large-Scale Nonstationary Spatial Covariance Function Estimation Using Convolutional Neural Networks},
journal = {Journal of Computational and Graphical Statistics},
volume = {34},
number = {2},
pages = {683--696},
year = {2025},
publisher = {ASA Website},
doi = {10.1080/10618600.2024.2402277},


URL = { 
    
        https://doi.org/10.1080/10618600.2024.2402277
    
    

},
eprint = { 
    
        https://doi.org/10.1080/10618600.2024.2402277
    
    

}
}

@article{good2013,
  title={{EN4}: Quality controlled ocean temperature and salinity profiles and monthly objective analyses with uncertainty estimates},
  author={Good, Simon A and Martin, Matthew J and Rayner, Nick A},
  journal={Journal of Geophysical Research: Oceans},
  volume={118},
  number={12},
  pages={6704--6716},
  year={2013},
  publisher={Wiley Online Library}
}

@article{hosoda2008,
  title={A monthly mean dataset of global oceanic temperature and salinity derived from {A}rgo float observations},
  author={Hosoda, Shigeki and Ohira, Tsuyoshi and Nakamura, Tomoaki},
  journal={JAMSTEC Report of Research and Development},
  volume={8},
  pages={47--59},
  year={2008},
  publisher={Japan Agency for Marine-Earth Science and Technology}
}

@ARTICLE{argofloats,

AUTHOR={Kelley, Dan E.  and Harbin, Jaimie  and Richards, Clark },

TITLE={{argoFloats}: An {R} Package for Analyzing {A}rgo Data},

JOURNAL={Frontiers in Marine Science},

VOLUME={8},

YEAR={2021},

pages = {635922},

URL={https://www.frontiersin.org/journals/marine-science/articles/10.3389/fmars.2021.635922},

DOI={10.3389/fmars.2021.635922},

ISSN={2296-7745},
}

@article{vu2022modeling,
  title={Modeling nonstationary and asymmetric multivariate spatial covariances via deformations},
  author={Vu, Quan and Zammit-Mangion, Andrew and Cressie, Noel},
  journal={Statistica Sinica},
  volume={32},
  number={4},
  pages={2071--2093},
  year={2022},
  publisher={JSTOR}
}
\end{document}


\renewcommand{\thesection}{S.\arabic{section}}
\renewcommand{\thesubsection}{S.\arabic{section}.\arabic{subsection}}
\renewcommand{\thesubsubsection}{S.\arabic{section}.\arabic{subsection}.\arabic{subsubsection}}
\numberwithin{equation}{section}
\renewcommand{\theequation}{S.\arabic{section}.\arabic{equation}}
\renewcommand{\thefigure}{S.\arabic{figure}}
\renewcommand{\thetable}{S.\arabic{table}}
\maketitle


\section{Application to vegetation index data}
\label{sec:ndvi_application}

In this section we apply the proposed GP$_{\mathrm{NAF}}$ model to a large dataset, specifically the Normalized Difference Vegetation Index (NDVI) data covering the North Slope of Alaska, made available by \citet{rocha2018arctic}. NDVI is a remotely sensed measure of vegetation activity and has been widely used to study changes in Arctic vegetation and the ``Arctic greening'' phenomenon. The dataset is part of the Arctic-Boreal Vulnerability Experiment (ABoVE) and contains growing-season (June--August) average NDVI at a spatial resolution of 1~km for each year from 2000 to 2015. The observations are derived from the Moderate Resolution Imaging Spectroradiometer (MODIS) MOD13A2 product \citep{rocha2019modis}. For our analysis, we consider 50,000 observations selected at random from those in the year 2002.  Each row in the data set corresponds to a $1~\mathrm{km}^2$ grid cell containing latitude and longitude coordinates, and NDVI \citep{rocha2019modis}. Prior to model fitting, the NDVI values were standardized by subtracting the empirical mean and dividing by the empirical standard deviation of all available data in 2002.

\begin{figure}[h!]
    \centering
    \begin{tabular}{cc}
        \textbf{Observed NDVI} &
        \textbf{GP$_{\mathrm{NAF}}$ prediction} \\[1mm]
        \includegraphics[width=0.48\textwidth]{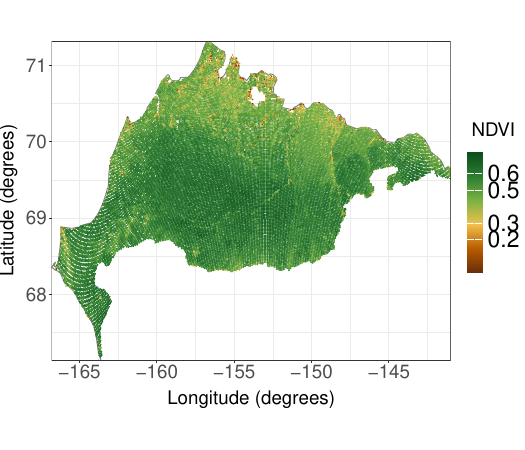} &
        \includegraphics[width=0.48\textwidth]{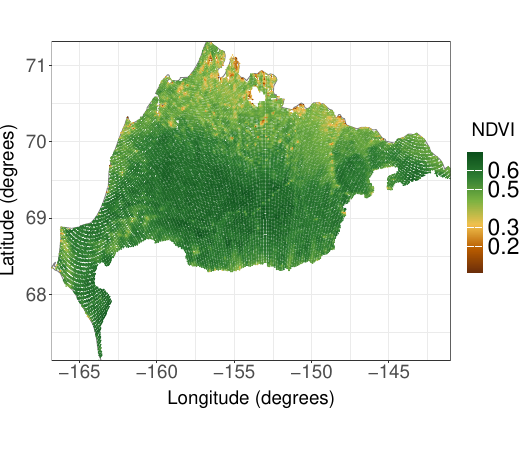} \\[3mm]
        & \textbf{GP$_{\mathrm{NAF}}$ prediction standard error} \\[1mm]
        &     \includegraphics[width=0.48\textwidth]{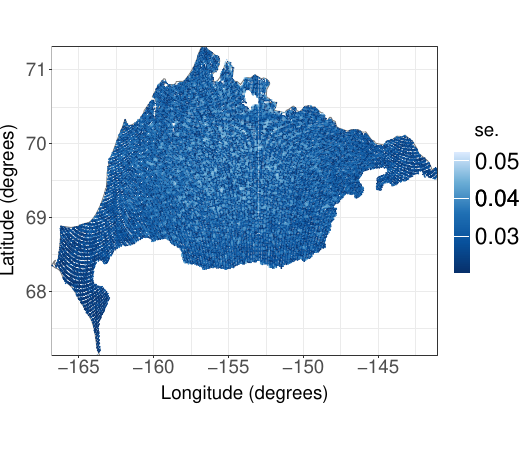}
    \end{tabular}
    \caption{Spatial prediction for the NDVI data. Top left: Observed NDVI. Top right: prediction from GP$_{\mathrm{NAF}}$. Bottom right: prediction standard error from GP$_{\mathrm{NAF}}$.}
    \label{fig:ndvi_prediction}
\end{figure}

To deal with the large number of spatial locations with GP$_{\mathrm{NAF}}$, we apply a maximin ordering \citep{guinness2018permutation} to the spatial locations and then implement a nearest-neighbor approximation on the ordered data as described in \citet{katzfuss2021general}. We find the maximum likelihood estimates of the GP$_{\mathrm{NAF}}$ parameters using minibatches. This combination of the Vecchia approximation and minibatch optimization substantially reduces the computational cost associated with the large number of spatial locations. We compare GP$_{\mathrm{NAF}}$ to a version of GP$_{\mathrm{deepspat}}$ that also uses a nearest-neighbour approximation \citep{Vu_2025}, using the same architecture as in the main text for both models. Table~\ref{tab:ndvi_comparison} summarizes the predictive performance of the two models on all the 174,343 data points in 2002 using the mean squared error (MSE), prediction interval coverage probability (PICP), and mean prediction interval width (MPIW). As can be seen, GP$_{\mathrm{NAF}}$ outperforms GP$_{\mathrm{deepspat}}$ in terms of predictive accuracy, as shown by the lower MSPE value. The PICP values for both models are close to the nominal level of 0.95, indicating that both models provide well-calibrated prediction intervals. However, GP$_{\mathrm{NAF}}$ achieves this with a narrower MPIW, and is hence more statistically efficient. In terms of computational efficiency, GP$_{\mathrm{NAF}}$ required similar compute time to GP$_{\mathrm{deepspat}}$, taking 79 minutes to fit compared to 93 minutes for GP$_{\mathrm{deepspat}}$. Overall, these results demonstrate the effectiveness of GP$_{\mathrm{NAF}}$ in modeling nonstationary spatial processes and its potential for large-scale spatial prediction tasks. Figure~\ref{fig:ndvi_prediction} shows the data together with the spatial predictions and prediction standard errors obtained from GP$_{\mathrm{NAF}}$.

\begin{table}[t]
    \centering
    \caption{Predictive performance for the NDVI data. Smaller MSPE values indicate better predictive performance.}
    \label{tab:ndvi_comparison}
    \begin{tabular}{|lcccc|}
        \toprule
        Model & MSPE & PICP & MPIW & Time (in mins)\\
        \midrule
        GP$_{\mathrm{deepspat}}$ & 0.0021 & 0.96 & 0.22 & 93\\
        GP$_{\mathrm{NAF}}$     & 0.0008 & 0.94 & 0.16 & 79\\
        \bottomrule
    \end{tabular}
\end{table}

\begin{figure}[t!]
    \centering
    \includegraphics[]{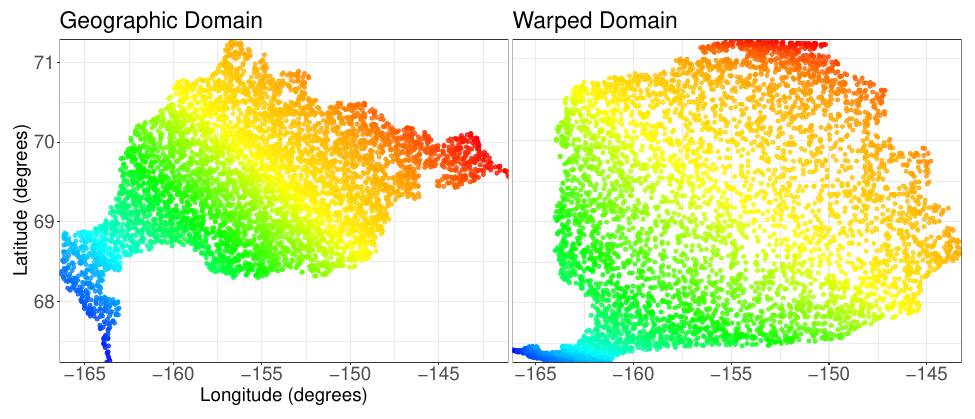}
    \caption{Learned spatial warping of the North Slope of Alaska. The left panel shows points in the geographic domain with colors varying uniformly by latitude and longitude, while the right panel shows the corresponding points in the warped domain, as inferred using GP$_{\text{NAF}}$.}
    \label{fig:ndvi_warping}
\end{figure}

The learned spatial deformation provides a means of examining how GP$_{\mathrm{NAF}}$ captures nonstationarity in the NDVI field. Figure~\ref{fig:ndvi_warping} compares the spatial coordinates in the original geographic domain with those obtained after applying the learned deformation. The observed NDVI field (Figure~\ref{fig:ndvi_prediction}) exhibits distinct linear spatial patterns, particularly in regions associated with mountainous terrain, suggesting substantial spatial heterogeneity and potentially varying dependence structures across the domain. The learned warping adapts to these features by stretching regions in the central part of the domain while compressing regions towards the boundaries. This deformation redistributes the spatial locations in such a way that the NDVI is well-approximated by a statoinary process in the transformed domain.

\bibliographystyle{apalike}

\bibliography{ref}